\documentclass{article}

\usepackage{amsmath}

\numberwithin{equation}{section}

\usepackage{amsfonts}

\usepackage{amsfonts}

\usepackage{slashed}

\def\be{\begin{equation}}
\def\ee{\end{equation}}
\def\bea{\begin{eqnarray}}
\def\eea{\end{eqnarray}}
\def\({\left(}
\def\){\right)}
\def\<{\left<}
\def\>{\right>}

\def\|{\left|}
\def\!{\right|}
\def\tr{{\mbox{tr}}}
\def\be{\begin{equation}}
\def\ee{\end{equation}}
\def\bea{\begin{eqnarray*}}
\def\eea{\end{eqnarray*}}
\def\ben{\begin{eqnarray}}
\def\een{\end{eqnarray}}
\def\({\left(}
\def\){\right)}
\def\<{\left<}
\def\>{\right>}

\def\[{\left[}
\def\]{\right]}

\def\x{\right|}
\def\+{\bar}
\def\mb{\mathbb}
\def\tr{{\mbox{tr}}}

\def\D{{\cal{D}}}
\def\L{{\cal{L}}}
\def\H{{\cal{H}}}
\def\K{{\cal{K}}}

\def\p{{\vec{p}}}
\def\q{{\vec{q}}}
\def\x{{\vec{x}}}

\def\t{\tilde}

\def\t{\widetilde}

\def\Z{{\cal{Z}}}

\def\G{{\cal{G}}}

\def\ee{\breve{e}}

\def\+{\breve{+}}
\def\-{\breve{-}}

\def\G{{\cal{G}}}

\def\Vol{\mbox{Vol}}

\begin{document}

\setlength{\unitlength}{1mm}

\pagestyle{empty}
\vskip-10pt
\vskip-10pt
\hfill 
\begin{center}
\vskip 3truecm
{\Large \bf
M5/D4 brane partition function}\\
\vskip0.3cm
{\Large \bf  on a circle bundle}\\
\vskip 2truecm
{\large \bf
Dongsu Bak\footnote{dsbak@uos.ac.kr}, Andreas Gustavsson\footnote{agbrev@gmail.com} }\\
\vskip 1truecm
{\it  Physics Department, University of Seoul, 13 Siripdae, Seoul 130-743 Korea}
\end{center}
\vskip 2truecm
{\abstract{We consider Abelian M5 brane on a six-manifold which we take as a circle bundle over a five-manifold $M$. We compute the zero mode part of the M5 brane partition function using Chern-Simons theory and Hamiltonian formulation respectively and find an agreement. We also show that the D4 brane on $M$ shares exactly the same zero mode partition function again using the Hamiltonian formulation. For the oscillator modes we find that KK modes associated with the circle compactification are missing from the D4 brane. By making an infinitesimal noncommutative deformation we have instanton threshold bound states. We explicitly compute the instanton partition function up to instanton charge three, and show a perfect match with a corresponding contribution inside the M5 brane partition function, thus providing a very strong supporting evidence that D4 brane is identical with M5 brane which extends beyond the BPS sector. We comment on the modular properties of the M5 brane partition function when compactified on $T^2$ times a four-manifold. We briefly discuss a case of a singular fibration.}}

\vfill
\vskip4pt
\eject
\pagestyle{plain}

\section{Introduction}
On the M5 brane worldvolume $W$ there lives a selfdual two-form $B$,  whose field strength $H=dB$ is selfdual. There is no covariant action one can write down for a selfdual two-form (without including auxiliary fields) because $H\wedge H = 0$. If one computes the partition function of a non-selfdual two-form, one finds for the zero mode part a sum of terms
\bea
\Z_{zero} &=& \sum_{a,b} \Theta\[^{a}_{b}\](\tau) \overline{\Theta\[^{a}_{b}\](\tau)}
\eea
Here $\tau$ is the period matrix on the intermediate Jacobian
\bea
J_W &=& H^3(W,\mb{R})/H^3(W,\mb{Z})
\eea
and $a^i,b_i$ are characteristics each of them taking values $0$ or $\frac{1}{2}$, and $i=1,\cdots,\frac{b_3}{2}$ where $b_3=$ dim$H^3(W,\mb{R})$.  $\Theta\[^{a}_{b}\](\tau)$ denotes the Jacobi theta function, and one of these theta functions will correspond to the partition function of a selfdual two-form on $W$ embedded in some eleven-manifold. In general it appears the M5 brane partition function is not completely diffeomorphism invariant. To see this, let us take the worldvolume as $W = T^2 \times \mb{CP}^2$  \cite{Witten:1996hc}\footnote{One may object that $\mb{CP}^2$ has no spin structure. However the fermions on the M5 brane also carry an R-symmetry index and such fermions may live on $T^2 \times \mb{CP}^2$. It is the eleven-dimensional space which must be a spin manifold, rather than the M5 brane worldvolume. One may also consider Taub-NUT space $TN$ in place of $\mb{CP}^2$ which also has a unique normalizable selfdual harmonic two-form. However this space is noncompact and so it might be less clear how to define the full partition function.}. We have precisely one selfdual harmonic two-form on $\mb{CP}^2$ and the theory of selfdual two-form reduces to the theory of a compact chiral scalar field on $T^2$. It is well-known that no modular invariant partition function exists for a compact chiral scalar on $T^2$. What we can require is the M5 brane partition function be invariant only under diffeomorphisms that preserve the physical data which determine its characteristics. Such physical data may consist of the orientation of $W$ and the spin structure of $W$ \cite{Witten:1996hc}. How the spin structure dependence on the characteristics appears is complicated and we will not address the question in any satisfactory way. It appears that the framework where this question can be answered is within seven-dimensional Chern-Simons theory. In the first part of this paper we follow \cite{Witten:1996hc} to compute the partition function of a selfdual two-form on a generic six-manifold from Chern-Simons theory, but leave the determination of the characteristics aside perhaps for the future.

We next perform a direct computation of the M5 brane partition function starting from the M5 brane action. Previous works on direct computations include \cite{Dolan:1998qk}, \cite{Henningson:1999dm}, \cite{Henningson:2001wh}. The separation of the partition function into a zero mode part and an oscillator mode part is somewhat intricate. We can illustrate this by taking a real compact 2d scalar on $T^2$. Its full modular invariant partition function is given by
\bea
\Z &=& \sum_{a,b} \left|\frac{\Theta\[^a_b\](\tau)}{\eta(\tau)}\right|^2
\eea
Here $\eta(\tau)$ denotes the Dedekind eta function. If we use the Hamiltonian formulation, then we are led to the following separation of this partition function into zero mode and oscillator mode parts, \cite{Polchinski:1998rq}
\bea
\Z_{zero} &=& \sum_{a,b} \left|\Theta\[^a_b\](\tau)\right|^2\cr
\Z_{osc} &=& \frac{1}{|\eta(\tau)|^2}
\eea
neither of which are modular invariant by themselves, and only the product of these is modular invariant. If on the other hand we use the Lagrangian formulation, then it is not so obvious how we shall define the measure in the path integral. One way is to define the measure so that we end up with the following separation \cite{Ginsparg:1988ui}
\bea
\Z_{zero} &=& \tau_2 \sum_{a,b} \left|\Theta\[^a_b\](\tau)\right|^2\cr
\Z_{osc} &=& \frac{1}{\tau_2|\eta(\tau)|^2}
\eea
again neither of which are modular invariant by themselves\footnote{A modular invariant combination is $\sqrt{\tau_2} |\eta(\tau)|^2$}.

One of our goals in this paper is to establish an equivalence between M5 brane partition function on a circle-bundle over some five-manifold $M_5$, and a corresponding partition function for D4 brane on $M_5$ obtained by dimensional reduction along the fiber. To this end we need to preserve the separation between zero modes and oscillator modes in our comparison between M5 and D4 brane. This means that we can not work in the Lagrangian formulation to obtain the zero mode contribution of the D4 brane partition function as was done recently in \cite{Gustavsson:2011ur}, \cite{Dolan:2012fu}, and try to match this zero mode contribution with the zero mode contribution of the M5 brane when that is computed using the Hamiltonian formulation. In fact a mismatch is shown explicitly to arise between zero modes for D4 and M5 brane by using this approach in \cite{Dolan:2012fu}. But this mismatch is entirely due to the fact that different formulations are used to compute zero modes. By using the Hamiltonian formulation for both M5 and D4 brane, we will see that their zero mode contributions respectively perfectly agree.

As we already indicated above, a careful treatment of zero mode and oscillator mode contributions is also crucial for modular invariance. In \cite{Dolan:1998qk} it was shown that for the selfdual tensor field part of M5 brane on $T^6$ this gives rise to an $SL(6,\mb{Z})$ modular invariant partition function for the choice of $00$-characteristics. For this we need to combine zero-mode and oscillator-mode contributions. In this paper we also show that the scalar fields part alone is $SL(6,\mb{Z})$ invariant by combining zero mode and oscillator mode contributions. For generic six manifolds of the form $T^2\times M_4$ we study $SL(2,\mb{Z})$ mapping class group acting on the $T^2$. Here we show that the M5 brane partition function with $00$-characteristic is modular invariant up to a phase.

Considering now the oscillator modes of D4 on $M_5$, we find that all KK-modes along the
circle direction are missing. Thus, at this stage, we find a  mismatch of the partition functions
 in their oscillator parts. But of course this is not the end of the story. Taking $S^1$ as an M-theory circle direction, a KK momentum along the circle direction is interpreted as
a D0 brane in the type IIA string theory side. When D0 branes are bound to D4, their dynamics can be faithfully described by worldvolume degrees of freedom on the D4 brane.  These are nothing but instantonic particles which satisfy the (anti) selfdual equation
along the four spatial directions of D4. For our $U(1)$ case, however, the corresponding instanton solutions become singular and, consequently, missed when one considers only regular configurations. In order to regularize them, we introduce a spatial noncommutativity which makes the size of instanton finite
and, then, take the commutative limit in the end recovering the original symmetries of D4 brane.

One may wonder whether the D0 brane could escape the D4 brane when it becomes singular in the commutative limit. However in \cite{Witten:1995zh} it was argued that the Higgs and Coulomb branches are decoupled even when the instanton shrinks to zero size. This argument was further supported by an index calculation made in \cite{Sethi:1996kj} for instanton number one. \footnote{We would like to thank Sethi for pointing out this problem to us, as well as informing us about his reference.}  

In this paper we show that these small instanton contributions match precisely with those of the missing KK modes,
which is in accordance with  the original idea of the proposal in \cite{Lambert:2010iw,
Douglas:2010iu}. In particular
one can be rather explicit for the one-instanton case. For  two instantons,
in addition to the Hilbert space of two separate instantons,
one can find so-called threshold bound state in
their moduli space which corresponds to a single $p_5=2$ KK mode of the M5 side \cite{Lee:2000hp}.
This counting  can be continued to higher instanton numbers \cite{Kim:2011mv}.

Finally we discuss the case of singular fibration in which the circle size $R$ becomes zero at some points of base
manifold \cite{Witten:2009at}. For this, in addition to the usual 5d sYM fields on D4 worldvolume, one may need some extra degrees localized
at the singular points in order to have a full agreement.  We illustrate this phenomenon rather explicitly in the example of
$TN\times T^2$ where one has a codimension-four singularity if one takes the M-theory circle direction to be that of the
Taub-NUT
circle. We show that the extra degree needed is a chiral 2d scalar on $T^2$.

Other works which are of some relevance to various aspects of our work include \cite{Hatefi:2012sy}, \cite{Tan:2008wp}.

In Appendix $A$, we obtain the period matrix for six-manifolds $W$ which are such types of circle-bundles in which the fiber-circle constitutes a one-cycle in $W$. So for instance this excludes the case of $W = S^3 \times S^3$. Even though $S^3$ is a circle-bundle there is no one-cycle in $S^3$. In Appendix $B$ we obtain the period matrix for $W = T^2\times M_4$ where for simplicity we assume there are no one-cycles on $M_4$. In Appendix $C$ we present holomorphic factorization of the partition function of a nonchiral boson (applicable to  2d scalar and 6d two-form) at the free fermion radius.

\section{M5 brane from Chern-Simons theory}
We follow \cite{Witten:1996hc} and consider the action of a non-chiral two-form $B$ such that only its selfdual part couples to a background three-form gauge potential $C$ whose field strength is $G=dC$,
\ben
S[B,C] &=& -\frac{\lambda}{2} \int_W \(|H|^2 + 2B \wedge G\)\label{M5}
\een
Here $\lambda$ is a coupling constant, $W$ denotes the six-dimensional world-volume of the M5 brane, $|H|^2 = H \wedge *H$ where we define $H=dB + C$. Our definition of Hodge dual is found in appendix $A$. This action is the six-dimensional analog of a gauged WZW action on a two-manifold to which there are many similarities. The action is not invariant under the gauge symmetry $\delta C = d\Lambda$, $\delta B = -\Lambda$ but it transforms as
\bea
\delta S[B,C] &=& \lambda \int_W \Lambda \wedge G
\eea
As shown in  \cite{Witten:1996hc} this gauge anomaly is canceled by the term $\sim \int C\wedge G\wedge G$ in the M-theory effective action. For our purpose, the gauge anomaly will be helpful in order to match with Chern-Simons theory in seven dimensions, so for the moment we like to keep it, rather than canceling it. From the action, we derive the equation of motion $d * H = G$. We also have the Bianchi identity $dH = G$. On $W$ we may decompose any three-form $\omega$ as $\omega = \omega^+ + \omega^-$ where $*\omega^+ = \pm \omega^{\pm}$. The equation of motion and the Bianchi identity are consistent with the selfduality equation $H^- = 0$ and it is in this sense that the action describes a selfdual three-form $H^+$.  Classically this selfduality equation is consistent for any value of $\lambda$. But as we will see, the situation changes in the quantum theory. If we introduce the inner product $(\omega,\eta) = \int_W \omega \wedge * \eta$ then we can write the M5 brane action in the form
\bea
S[B,C] &=& -\frac{\lambda}{2} \( (dB,dB) + 2(C^+,C^-) + 4(dB,C^-)\)
\eea
Let us define the partition function of $B$ in the background of $C$ as
\bea
\Z(C) &=& \int \D B e^{iS[B,C]}
\eea
By differentiating under the integral sign, we then find that
\bea
\(\frac{\delta}{\delta C^+} + i \lambda C^-\) \Z(C) &=& 0
\eea
and
\bea
\(d \frac{D}{DC^-} + 2i \lambda G\) \Z(C) &=& 0
\eea
In the second relation we assume the equation of motion $dH^+ = G$ and we define a covariant derivatives as
\bea
\frac{D}{DC^-} &=& \frac{\delta}{\delta C^-} - i\lambda C^+\cr
\frac{D}{DC^+} &=& \frac{\delta}{\delta C^+} + i\lambda C^-
\eea
Our functional derivatives are defined with respect to the inner product as $\frac{\delta}{\delta \omega} (\omega,\eta) = \eta$.

Let us now consider Chern-Simons theory at level $k$ on a seven-manifold $U$ bounded by $W$
\bea
S &=& \frac{k}{4\pi} \int_U C \wedge dC
\eea
As usual we can make a variation and read off the symplectic potential from the boundary term. If we assume that $\partial U = W$, then we have the symplectic potential
\ben
A &=& -\frac{k}{4\pi} \int_W C \wedge \delta C\label{sympl1}
\een
which we can also write as
\bea
A &=& \frac{k}{4\pi} \(-(C^-,\delta C^+) + (C^+,\delta C^-)\)
\eea
Here we have noted that the inner product of two selfdual or two antiselfdual three-forms vanish identically. Its curvature is the symplectic two-form
\bea
\Omega &=& \frac{k}{2\pi} (\delta C^+,\delta C^-)
\eea
In components we have
\bea
\Omega_{+-} &=& \frac{k}{2\pi}
\eea
and the inverse is
\bea
\Omega^{-+} &=& \frac{2\pi}{k}
\eea
We define covariant derivatives
\bea
\frac{D}{DC^{\pm}} &=& \frac{\delta}{\delta C^{\pm}} - i A_{C^{\pm}}
\eea
where
\bea
A_{C^{\pm}} &=& \mp\frac{k}{4\pi} C^{\mp}
\eea
consistent with
\bea
\lambda &=& \frac{k}{4\pi}
\eea
Gauge transformations act on a wave function $\psi(C)$ and the gauge potential as $\psi \rightarrow  e^{i\Lambda} \psi$, $A_{C^{\pm}} \rightarrow  A_{C^{\pm}} + \frac{\delta \Lambda}{\delta C^{\pm}}$. Under a gauge variation $\delta C = d\Lambda$ the Chern-Simons action varies by
\bea
\delta S &=& \frac{k}{4\pi} \int_W \Lambda \wedge G
\eea
This variation is identical with the gauge variation of the M5 brane action if we choose the coupling constant as $\lambda = \frac{k}{4\pi}$.

We impose holomorphic polarization of the wave function $\psi$,
\bea
\frac{D}{DC^+} \psi &=& 0
\eea
In temporal gauge we have the Gauss law constraint $d(C^+ + C^-) = 0$. After quantization this becomes $d \(P(C^+) + P(C^-)\) \psi = 0$ where the prequantum operators are given by
\bea
P(C^+) &=& -\frac{2\pi i}{k} \frac{D}{D C^-} + C^+\cr
P(C^-) &=& \frac{2\pi i}{k} \frac{D}{D C^+} + C^-
\eea
The general solution to the holomorphic polarization condition is of the form
\bea
\psi(C^+,C^-) &=& \exp \(\frac{ik}{4\pi} (C^+,C^-)\) \phi(C^-)
\eea
and so
\bea
P(C^+) &=& -\frac{2\pi i}{k} D_- + C^+\cr
P(C^-) &=& C^-
\eea
From this we get the quantum version (\ref{Gauss}) of Gauss law as
\ben
\(d \frac{D}{DC^-} + \frac{ik}{2\pi} G\) \psi &=& 0\label{Gauss}
\een
We see the same equations from the M5 brane action if we choose the coupling $\lambda=\frac{k}{4\pi}$. This implies that we shall identify the wave function $\psi$ with the M5 brane partition function $\Z$.

\subsection{Zero mode part}
Let us now compute the zero mode part of the wave function $\psi$. We expand the harmonic part of $C$ in a basis of harmonic three-forms on $W$. We can choose this as a symplectic basis. If we let $a_i$ and $b^i$ denote three-cycles in $W$, then $\alpha_i$ and $\beta^i$ which denote the Poincar\'e dual harmonic three-forms which we normalize as
\bea
\frac{1}{2\pi} \int_{a_j} \alpha_i &=& \delta_i^j\cr
\frac{1}{2\pi} \int_{b^j} \beta^i &=& \delta^i_j
\eea
will constitute a symplectic basis on $W$ in the sense that
\bea
\(\begin{array}{cc}
\int_W \alpha_i \wedge \alpha_j & \int_W \alpha_i \wedge \beta^l \\
\int_W \beta^k \wedge \alpha_j & \int_W \beta^k \wedge \beta^l
\end{array}\)
&=& 4\pi^2 \(\begin{array}{cc}
0 & \delta_i^l\\
-\delta^k_j & 0
\end{array}\)
\eea
We will also define the following linear combinations
\bea
\omega_i &=& \alpha_i + \tau_{ij} \beta^j\cr
\bar{\omega}_i &=& \alpha_i + \bar{\tau}_{ij} \beta^j
\eea
by demanding these are selfdual and antiselfdual harmonic three-forms respectively. The period matrix $\tau_{ij}$ can then be extracted by integrating $\omega_i$ over the $b^j$-cycles,
\bea
\frac{1}{2\pi} \int_{b^j} \omega_i &=& \tau_{ij}
\eea
If we define
\bea
\tau &=& -\tau_1 - \tau_2\cr
\bar{\tau} &=& -\tau_1 + \tau_2
\eea
then we have
\bea
\int_W \omega_i \wedge \bar{\omega}_j &=& 8\pi^2 (\tau_2)_{ij}
\eea
We expand the harmonic part of the $C$-field as
\bea
C &=& x^i \alpha_i + y_i \beta^i
\eea
In terms of complex coordinates that we define as
\bea
a_i &=& \bar{\tau}_{ij} x^j - y_i\cr
\bar{a}_i &=& \tau_{ij} x^j - y_i
\eea
we have
\bea
C &=& \frac{1}{2} (\tau_2)^{ij} \(a_i \omega_j - \bar{a}_i \bar{\omega}_j\)
\eea
Large gauge transformations leave the holonomy
\bea
H(C) &=& e^{i \int_C C}
\eea
invariant for any choice of three-cycle $C$. Since the connection is constrained to be flat by the Chern-Simons action, the exponent defines an element in $J_W$ which is a torus $T^{b_3(W)}$ whose coordinates are $(x^i,y_j)$ subject to torus identifications
\bea
x^i &\sim & x^i + 1\cr
y_j &\sim & y_j +1
\eea
Large gauge variations act on $J_W$ as
\ben
\delta x^i &=& n^i\cr
\delta y_j &=& m_j\label{large}
\een
for integers $n^i,m_j$. The Chern-Simons Lagrangian becomes\footnote{While a non-Abelian generalization of the M5 brane Lagrangian has not been found, one can probably more easily write down a non-Abelian generalization of the reduced Chern-Simons quantum mechanics Lagrangian on the space of zero modes (the intermediate Jacobian). Here on the intermediate Jacobian we have just complex-valued scalar fields $a_i$ which depend only on time.}
\bea
L &=& \frac{k\pi}{2} (\tau_2)^{ij} \(-a_i \dot{\bar{a}}_j + \bar{a}_i \dot{a}_j\)
\eea
and we get the symplectic potential and symplectic two-form as
\ben
A &=& \frac{k\pi}{2} (\tau_2)^{ij} \(-a_i \delta \bar{a}_j + \bar{a}_i \delta a_j\)\label{original}\\
\Omega &=& -k\pi (\tau_2)^{ij} \delta a_i \wedge \delta \bar{a}_j
\een
Canonical commutation relations are
\ben
[a_i,\bar{a}_j] &=& \frac{i}{k\pi} (\tau_2)_{ij}\label{cancomm}
\een
and prequantum operators are
\bea
P(a_i) &=& \frac{i}{k\pi} (\tau_2)_{ij} D_{\bar{a}_j} + a_i\cr
P(\bar{a}_i) &=& -\frac{i}{k\pi} (\tau_2)_{ij} D_{a_j} + \bar{a}_i
\eea
which one may check realize the algebra (\ref{cancomm}).

Large gauge transformations act on the potential and the wave function as
\bea
A' &=& A + \delta \Lambda(n,m)\cr
\psi' &=& e^{i \Lambda(n,m)} \psi
\eea
where we define
\bea
A'(a,\bar{a}) &=& A(a+\bar{\tau} n - m,a+\tau n - m)\cr
\psi'(a,\bar{a}) &=& \psi(a+\bar{\tau} n - m,a+\tau n - m)
\eea
For the sake of convenience (and not for pedagogical reasons), we will switch back and forth between real and complex coordinates. As we want to see how various quantities depend on the various choices we make, we will now assume the symplectic potential is chosen as
\bea
A &=& 2\pi k x^i \delta y_i + \delta \mu(x^i,y_j)
\eea
with an arbitrary function $\mu$. Such a gauge transformation of the symplectic potential corresponds to a canonical transformation of the phase space variables $x^i,y_j$. We impose holomorphic polarization on the wave function
\bea
D_{a_i} \psi(a,\bar{a}) &=& 0
\eea
This condition is solved by
\bea
\psi(a,\bar{a}) &=& e^{i\(K(a,\bar{a})+\mu(a,\bar{a})\)} \phi(\bar{a})
\eea
where
\bea
K &=& \pi k \tau_{ij} x^i x^j + f(\bar{a})
\eea
is the Kahler potential. In the Kahler potential we have the freedom of adding an arbitrary antiholomorphic function $f$.

We can read off the gauge parameter associated to a large gauge transformation (\ref{large}) from the variation of the symplectic potential
\bea
\delta A = A(x+n,y+m) - A(x,y) = \delta \Lambda(n,m)
\eea
This gives us the gauge parameter as
\bea
\Lambda(n,m) &=& 2\pi k n^i y_i + \mu(x+n,y+m) - \mu(x,y) + c(n,m)
\eea
where $c(n,m)$ is a further closed term that we can always add. From
\bea
e^{i\Lambda(n,m)} \psi(x,y) &=& \psi(x+n,y+m)
\eea
we find that $\mu$ cancels out and we are left with the condition that
\bea
e^{2\pi i k n^i y_i} e^{i c(n,m)} e^{iK(a,\bar{a})} \phi(\bar{a}) &=& e^{iK(a+\delta a,\bar{a}+\delta\bar{a})}\phi(\bar{a} + \delta \bar{a})
\eea
Here
\bea
K(a+\delta a,\bar{a}+\delta\bar{a}) -K(a,\bar{a}) &=& \pi k \tau_{ij} n^i n^j + 2\pi k \tau_{ij} n^i x^j + f(\bar{a}+\delta \bar{a}) - f(\bar{a})
\eea
and so we have, by noting that
\bea
n^i \(y_i - \tau_{ij} x^j\) &=& \bar{a}_i
\eea
 that
\bea
\phi(\bar{a}+\delta \bar{a}) &=& e^{i c(n,m)} e^{f(\bar{a}+\delta \bar{a}) - f(\bar{a})} e^{-i\pi k \tau_{ij} n^i n^j + 2\pi i k n^i \bar{a}_i} \phi(\bar{a})
\eea
Here
\bea
\delta \bar{a}_i &=& \tau_{ij} n^j - m_i
\eea
and $c(n,m)$ can be partially fixed by requiring that $\phi(\bar{a})$ satisfies the `cocycle condition' of large gauge transformations
\ben
\phi(\bar{a} + \tau (n+n') - m-m') &=& \phi((\bar{a}+\tau n' - m') + \tau n - m)\label{cocycle}
\een
Here, on the right-hand side our notation is supposed to mean that we first compute the gauge transformation $\phi(\bar{a}' + \tau n - m)$ in terms of $\phi(\bar{a}')$ and subsequently we express $\phi(\bar{a}')$ as a gauge transformation of $\phi(\bar{a})$. Let us first consider the case $f(\bar{a}) = 0$. We then find
\bea
c(n,m) &=& \pi k \tau_{ij} n^i m^j
\eea
is the minimal choice that makes the cocycle condition satisfied. But we can add further linear terms to this
\bea
c(n,m) &=&  \pi k \tau_{ij} n^i m^j  + 2\pi\( c_i n^i + d^i m_i\)
\eea
and still satisfy the cocycle condition. If we have non-vanishing $f$, we still get the same result for $c(n,m)$ due to the exponential nature of $e^{if}$ which automatically makes the cocycle condition satisfied.

If we choose $f=0$ then we can identify the gauge transformation of $\phi$ with the gauge transformation of a Jacobi theta function with certain characteristics, related to $c_i$ and $d^i$ as
\bea
\phi(\bar{a}) &=& \Theta\[^{\frac{c}{k}}_{-d}\](-k\tau|k\bar{a})
\eea

Let us now consider a three-cycle $C_{N,M}$ in $W$ characterized by periods
\bea
\int_{C_{N,M}} \omega_i &=& M_i + \tau_{ij} N^j
\eea
and define the holonomy
\bea
H(N,M) &=& e^{ik\int_{C_{N,M}} C}
\eea
over this three-cycle. When we insert the zero mode expansion for $C$ we get
\bea
H(N,M) &=& e^{i\pi k \((M+\tau N)_i (\tau_2)^{ij} a_j - (M+\bar{\tau}N)_i(\tau_2)^{ij}\bar{a}_j\)}\label{H-holonomy}
\eea
Since we do not make any canonical transformation of the phase space variable $C$ when we define the holonomy, we shall use the symplectic potential that we obtain as the boundary term of the Chern-Simons action, that is (\ref{original}). In this gauge we solve the polarization condition by
\bea
\psi(a,\bar{a}) &=& e^{-\frac{i\pi k}{2} (\tau_2)^{ij} a_i \bar{a}_j} \phi(\bar{a})
\eea
where $\phi(\bar{a})$ depends holomorphically on $\bar{a}$. The gauge variation of $\psi$ now induces the gauge variation
\ben
\phi(\bar{a} + \tau n - m) &=& e^{\frac{i\pi k}{2} (\tau_2)^{ij} \[(\bar{\tau}n-m)_i (\tau n-m)_j + 2 (\bar{\tau}n-m)_i \bar{a}_j\]} e^{i c(n,m)} \phi(\bar{a})\label{gaugetra}
\een
As before we can partially fix $c(n,m)$ by demanding that we satisfy the cocycle condition (\ref{cocycle}). We then again find that
\bea
c(n,m) &=& k\pi m_i n^i + 2\pi (c^i m_i + d_i n^i)
\eea
up to some constants $c^i$ and $d_i$ that we have to fix by other means. On the wave function $\phi$ the prequantum operators reduce to the quantum operators
\bea
Q(a_i) &=& -\frac{i}{\pi k} (\tau_2)_{ij} \partial_{\bar{a}_j} \cr
Q(\bar{a}_i) &=& \bar{a}_i
\eea
which also realize the algebra (\ref{cancomm}). We use the BCH formula
\bea
e^{A+B} &=& e^{-\frac{1}{2}[A,B]} e^A e^B
\eea
to express the holonomy in the form
\bea
H(N,M) &=& e^{-\frac{i\pi k}{2} (\tau_2)^{ij}(M+\bar{\tau}N)_i(M+\tau N)_j} e^{-i\pi k(M+\bar{\tau}N)_i (\tau_2)^{ij} \bar{a}_j} e^{(M+\tau N)_i \partial_{\bar{a}_i}}
\eea
When we act with the holonomy on $\phi(\bar{a})$ we then get
\bea
H(N,M) \phi(\bar{a}) &=& e^{-\frac{i\pi k}{2} (\tau_2)^{ij}(M+\bar{\tau}N)_i(M+\tau N)_j} e^{-i\pi k (M+\bar{\tau}N)_i (\tau_2)^{ij} \bar{a}_j} \phi(\bar{a} + M + \tau N)
\eea
Finally we use the gauge transformation of $\phi$ as given in (\ref{gaugetra}) and find that most exponential factors cancel out, and we are left with
\bea
H(N,M) \phi(\bar{a}) &=& e^{2\pi i (c^i M_i + d_i N^i)} e^{ik\pi M_i N^i} \phi(\bar{a})
\eea
We now see that the eigenvalue of the holonomy corresponds to the characteristics $c^i$ and $d_i$. Using the BCH formula we can show that the holonomies obey the algebra
\ben
H(N,M) H(N',M') &=& e^{\pi i k (M'_i N^i + M_i N'^i)} H(N+N',M+M')\label{algebra}
\een
We can be ignorant about the signs in the exponent since the difference is always of the form $e^{2\pi i k M_i N'^i} = 1$. For $k$ even this formula admits that we take all holonomies to be $H(N,M)=1$. For $k$ odd this formula forbids us to take all $H(N,M) = 1$ and the best we can do is to allow for some (or all) of them to be $-1$.

Let us now turn to holonomies on $J_W$,
\bea
W(N,M) &=& e^{i \int_{c_{N,M}} A}
\eea
As a preliminary attempt we define these by choosing the gauge potential as
\bea
A &=& 2\pi k x^i \delta y_i + 2\pi i (c^i \delta y_i + d_i \delta x^i)
\eea
since this makes $W(N,M)$ invariant under (\ref{large}) while any other choice obtained by adding for example adding an exact term on the form $\xi \delta (x^i y_i)$ would not give us an invariant holonomy under (\ref{large}) for generic values on $\xi$. We will return to this issue more fully below. Let us choose a closed loop on $J_W$ as a straight line
\bea
c_{N,M} &=& \{(x^i,y_j) = (N^i\theta,M_j\theta) | 0\leq \theta \leq 1\}\cr
&=& \{\bar{a}_i = (\tau_{ij}N^i - M_i)\theta  | 0\leq \theta \leq 1\}
\eea
The holonomy can now be evaluated to
\bea
W(N,M) &=& e^{2\pi i (c^i M_i + d_i N^i)} e^{ik\pi M_i N^i}
\eea
and we can read off the characteristics from
\bea
W(e^i,0) &=& e^{2\pi i c^i}\cr
W(0,e_j) &=& e^{2\pi i d_j}
\eea
where $e^i = (0,\cdots,1,\cdots,0)$ with $1$ at the $i$-th entry. However this does not bring us any closer to what the characteristics really should be.

Our definition of the holonomy $W(N,M)$ is not satisfactory since it is gauge choice dependent. We also can not see if using the definition of holonomy as proposed in \cite{Alvarez:1984es} can help us here. If we can find a gauge invariant definition of $W(N,M)$ we may expect that $W(N,M) = H(N,M)$ for that gauge invariant definition since clearly $H(N,M)$ is gauge invariant. Indeed there might exist a gauge invariant definition of $W(N,M)$. To allow for the most general possible extension of manifold and symplectic two-form, we shall include all the oscillator modes and we are led to consider seven-dimensional Chern-Simons theory to compute the holonomy following \cite{Witten:1996hc}. By including the oscillators, we have to consider the symplectic potential on the form $A = \frac{k}{4\pi} \int_W C \wedge \delta C$. We then consider a one-cycle $C$ in $J_W$ around which we want to compute the holonomy for this symplectic potential. Let us parameterize the one-cycle by $\theta \in [0,1]$ and consider a one-parameter family of $C$-fields that we denote as $C_C(\theta)$. Time coordinate is being replaced by $\theta$ coordinate. The differential $\delta C$ in field space becomes a differential in the seven dimensional space $S^1 \times W$ where $S^1$ is the one-cycle parametrized by $\theta$. The holonomy can be expressed as
\bea
H(C) = \exp i \int_C A = \exp \frac{ik}{4\pi} \int_{S^1 \times W} C_C \wedge dC_C
\eea
However this formula is only valid if $A$ is globally defined along $C$. For the generic case we instead define the holonomy by finding an extension $X$ whose boundary is $\partial X = S^1 \times W$ and define the holonomy as
\bea
H(C) = \exp \frac{ik}{4\pi} \int_X G \wedge G
\eea
where $G = dC_C$. This is still a simplification of the real situation in M theory since in general $G$ is an element in a shifted integral cohomology \cite{Witten:1996hc}. Nevertheless, the point is that we need to find an extension $X$ over which the spin structure of $S^1 \times W$ and the gauge field $C_C$ can be extended. How one can make such an extension may depend on the spin structure on $W$. For example, if we have periodic (Ramond) fermions on a circle we can not extend the circle to a disk. This means that the holonomy will depend on the spin structure on $W$.

One may consider $W = T^6$ as an example. Here $\Theta\[^0_0\]$ is the only fully modular invariant choice. But this need not necessarily be the partition function of the M5 brane of a given spin structure if that spin structure is not fully modular invariant. The M5 brane partition function need only be invariant under modular transformations that preserve the spin structure.

\subsection{Oscillators}
The inner product of two wave functions is defined as\footnote{In Euclidean signature we define the bar as usual complex conjugation. In this signature we also have $C^+ \wedge C^- = -i C^+ \wedge *C^-$ and $\overline{C^{\pm}} = C^{\pm}$. The exponent $\exp \frac{ik}{2\pi} (C^+,C^-)$ is now real with respect to our complex conjugation in Euclidean signature.}
\bea
(\psi,\psi) &=& \int \D C \overline{\psi(C)} \psi(C)\cr
&=& \int \D C^+ \D C^- \exp \(\frac{ik}{2\pi} (C^+,C^-)\) \overline{\phi(C^-)}\phi(C^-)
\eea
We can expand $C = C_{harmonic} + d\chi$ in a harmonic and an exact piece since by Gauss law $dC = 0$. We note that $\chi$ is real in both Euclidean as well as in Minkowski signature since Wick rotation can only affect the components $C_{0MN}$ which in our gauge choice are zero. The fact that $\chi$ is real means that it can be gauged away by a gauge transformation
\bea
\delta C &=& d\lambda
\eea
by taking the real gauge parameter as $\lambda = - \chi$. This is thus different from the quantization of three-dimensional Chern-Simons theory \cite{Bos:1989kn}. The generator of gauge transformations is given by the Gauss law constraint, which we here express for an infinitesimal gauge parameter $\lambda$ in the form
\bea
\int \lambda \wedge \(d\frac{\delta}{\delta C^-} - \frac{ik}{2\pi} G\) \phi(C^-) &=& 0
\eea
Upon integration by parts, this equation makes sense only if $\lambda$ is such that $(d\lambda)^+ = 0$ and $C^+ = 0$. In that case this equation reduces to
\bea
\phi(C^- + (d\lambda)^-) &=& \phi(C^-)
\eea
whose solution can be taken as
\bea
\phi(C^-) &=& \phi(C^-_{harmonic})
\eea
with a trivial oscillator contribution. The oscillator mode contribution to the inner product of wave functions is now given by
\bea
(\psi,\psi)_{osc} &=& \int \D C \delta(dC) \exp \(\frac{ik}{4\pi} \int |C|^2\)
\eea
where we have noticed that $(C^+,C^-) = \frac{1}{2} (C,C)$. We replace
\bea
\delta(dC) &=& \int \D \chi \exp \(\frac{ik}{2\pi}\int dC \wedge \chi\)
\eea
Completing the square and shifting $C \rightarrow C+d\chi$ and noting the measure is gauge invariant, $\D (C+d\chi) = \D C$, we get
\bea
(\psi,\psi)_{osc} &=& \int \D \chi \exp\(-\frac{ik}{4\pi} \exp |d\chi|^2\)
\eea
The path integral is suffering of a gauge redundancy $\delta \chi = d \lambda$ and so needs to be gauge fixed. It is clear that this is path integral is precisely equal to the path integral of a non-chiral two-form gauge field on $W$, but which here has been obtained from the Chern-Simons theory on $I \times W$.  The path integral was over $C$ but this is constrained to be flat $dC = 0$, and the exact part can therefore be re-expressed as an integral over $\chi$. We note that the sign gets correct. We express the path integral in Minkowski signature, and we have the phase factor $e^{iS}$ from which we read off the action
\bea
S &=& -\frac{k}{4\pi} \int |d\chi|^2
\eea
which corresponds to the action (\ref{M5}) in a vanishing background $C$ field if we make the identifications $B=\chi$.

\section{The M5 brane partition function}
In this section we perform a direct and explicit computation of the full Abelian M5 brane partition function using Hamiltonian formulation. For the selfdual three-form we recover the result we got above from Chern-Simons theory, but here we will also supplement this with contributions from the five scalar fields and the fermions.

In order to apply the Hamiltonian formulation we assume that the M5 brane worldvolume $M_6$ is a circle-bundle over a base-manifold $M_5$. We choose the following parametrization for the metric on $M_6$
\ben
ds^2_{M5} &=& \beta dt^2 + G_{mn} (dx^m - V^m dt)(dx^n - V^n dt)\label{inverse}
\een
and we assume that $\beta$ is a constant. We will assume that $\partial_t$ is a Killing vector field, which means that $\partial_t V^m = 0$ and $\partial_t G_{mn} = 0$. We will associate $t$ with time and $x^m$ with spatial directions on $M_5$. We will promote $\beta$ to a complex holomorphic parameter, thus allowing for Minkowskian and Euclidian signatures.

As the M5 brane action we will take
\bea
S &=& S_{B} + S_{\phi} + S_{\psi}
\eea
where
\bea
S_{B} &=& -\frac{\lambda}{12} \int d^6 x \sqrt{-g} H^{MNP}H_{MNP}\cr
S_{\phi} &=& -\lambda \int d^6 x \sqrt{-g} \partial^M \phi^A \partial_M \phi^A\cr
S_{\psi} &=& i \lambda \int d^6 x \sqrt{-g} \bar{\psi} \Gamma^M \partial_M \psi
\eea
Selfduality fixes the coupling constant to be
\bea
\lambda &=& \frac{1}{4\pi}
\eea

\subsection{Scalar field}\label{scalarzeromode}
Let us consider the scalar field action
\ben
S &=& -\lambda \int_{M_6} d^6 x \sqrt{-g} g^{MN} h_M h_N\label{scalar}
\een
for one of the five real scalar fields. Here we define the field strength of a zero form scalar field $\phi$ as
\bea
h_M &=& \partial_M \phi
\eea
The momentum conjugate $\phi$ is
\bea
\pi &=& -2\lambda \sqrt{-\beta} \sqrt{G} h^t
\eea
The Hamiltonian is
\bea
H &=& \lambda\sqrt{-\beta} \int_{M_5} d^5 x \sqrt{G} \(-h^t h_t + h_{6D}^m h_m\)
\eea
We note the following metric identities\footnote{Let us derive them here: $h_t = g_{tt} h^t + g_{tm} g^{mn} h_n + g^{tm} g^{mt} h_t$ gives $h_t(1-g_{tm}g^{mt}) = g_{tt} h^t + g_{tm} g^{mn} h_n$. We then miraculously notice that $g_{tt}$, $1-g_{tm}g^{mt}$ and $g_{tm}g^{mn}$ all contain the common factor $\beta + V^2$ which can be factored out and the identity follows. Next we compute $h^{mn}_{6D} = g^{mt} h_t + g^{mn} h_n = \frac{1}{\beta} V^m h_t + G^{mn} h_n + \frac{1}{\beta} V^m V^n h_n$ and substitute $h_t = g_{tt} h^t - V^n h_n$ and we find that two terms $\sim V^m V^n h_n$ cancel and the second identity follows.}
\bea
h_t &=& \beta h^t - V^m h_m\cr
h^m_{6D} &=& h^m + V^m h^t
\eea
and we see that the Hamiltonian can be expressed in terms of the five-dimensional metric as
\bea
H &=& \lambda\sqrt{-\beta} \int_{M_5} d^5 x \sqrt{G} \(-\beta (h^t)^2 + h^m h_m + 2 h^t V^m h_m\)
\eea
To get a better handle on the scalar field partition function, and in particular its modular invariance, we temporarily introduce a regulator and assume that the scalar field is compact
\bea
\phi(x) &\sim & \phi(x) + 2\pi r
\eea
with a radius $r$ that we in the end will take to infinity. The important point is that the regulator $r$ does not mix with the geometry of space-time, so we can introduce $r$ as a separate quantity on which the partition function may depend, in order to extract its dependence on the geometry. Using this together periodicity of the scalar field together with the canonical commutation relation
\bea
[\phi(x),\pi(y)] &=& i \delta^5(x-y)
\eea
we conclude that
\bea
P &=& - 2\lambda\sqrt{-\beta} \int d^5 y \sqrt{G} h^t
\eea
is integer quantized as
\bea
P &=& \frac{n}{r}
\eea
where $n$ is integer. Moreover, we expand $h^t$ in a basis of harmonic zero forms. But such harmonics are necessarily constants. So $h^t$ is a constant. It must therefore be given by
\bea
h^t &=& -\frac{n}{2\lambda\sqrt{-\beta} (\Vol) r}
\eea
where $\Vol = \int d^5 x \sqrt{G}$. The Hamiltonian is now given by
\bea
H &=& \frac{\lambda\sqrt{-\beta}}{\Vol} \frac{n^2}{r^2} +\cdots
\eea
where the $+\cdots$ terms will be irrelevant in the limit $r\rightarrow \infty$. In that limit the discrete sum can be replaced by an integral over $q = n/r$, and we get
\ben
\Z_{zero} = \int_{-\infty}^{\infty} dq e^{-2\pi i H} = \(\frac{\Vol}{2i\lambda\sqrt{-\beta}}\)^{\frac{1}{2}}\label{zero}
\een
Of course the integral is evaluated by taking $\beta$ real and positive, and then we make the analytic continuation.

\subsection{Tensor gauge field}
The tensor gauge field can be treated in an analogous way as the compact scalar field. The action for the non-chiral tensor gauge field is given by
\bea
S &=& -\frac{\lambda}{12} \int d^6 x \sqrt{-g} H^{MNP} H_{MNP}
\eea
We define the field strength of the two-form gauge potential $B_{MN}$ as
\bea
H_{MNP} &=& \partial_{M} B_{NP} + \partial_N B_{PM} + \partial_P B_{MN}
\eea
The momentum conjugate to $B_{MN}$ is
\bea
E^{MN} &=& -\frac{\lambda\sqrt{-\beta}}{2} \sqrt{G} H^{tMN}
\eea
and we have five primary constraints
\bea
E^{mt} &=& 0
\eea
for $m=1,\dots ,5$. These we must supplement by equally many gauge fixing constraints. We will choose the temporal gauge
\bea
B_{mt} &=& 0
\eea
The Hamiltonian is
\bea
H &=& \frac{\lambda \sqrt{-\beta}}{12} \int d^5 x \sqrt{G} \(-3 H^{tmn} H_{tmn} + H^{mnp} H_{mnp}\)
\eea
We note the following metric identities\footnote{The derivation is analogous to what we did for the scalar field.}
\bea
H_{0mn} &=& \beta H^t{}_{mn} - V^p H_{mnp}\cr
H^{mnp}_{6D} &=& H^{mnp} + 3 V^{[m} H^{|t|np]}
\eea
and we see that the Hamiltonian can be expressed in terms of the five-dimensional metric as
\bea
H &=& \frac{\lambda\sqrt{-\beta}}{12} \int d^5 x \sqrt{G} \(-3\beta H^{tmn} H^t{}_{mn} + 6 V^p H^{tmn} H_{mnp} + H_{mnp} H^{mnp}\)
\eea
The holonomies are periodic
\bea
\int_{\Sigma} B &\sim & \int_{\Sigma} B + 2\pi
\eea
Let us introduce the covariant momentum variable $e^{mn} = \frac{E^{mn}}{\sqrt{G}}$ whose indices we can covariantly lower by the five-dimensional metric to define a momentum two-form $e = \frac{1}{2} e_{mn} dx^m \wedge dx^n$. From the canonical commutation relations
\bea
[B_{mn}(x),E^{pq}(y)] &=& i \delta_{mn}^{pq} \delta^5(x-y)
\eea
we see that
\bea
\[\int_{\Sigma_j} B,\int_{B_5} e\wedge *\Omega^i\] &=& \frac{i}{2} \int_{\Sigma_j} \Omega^i
\eea
Here we denote by $\Omega^i$ the basis elements of harmonic two-forms on $M_5$ and we have a metric
\bea
\G^{ij} &=& \int_{M_5} \Omega^i \wedge * \Omega^j
\eea
If we expand
\bea
e &=& e_i \Omega^i
\eea
then we get
\bea
\[\int_{\Sigma_j} B,e^i\] &=& \frac{i}{2} \delta^i_j
\eea
where we define $e^i = \G^{ij} e_j$. Then
\bea
2e^i &=& n^i
\eea
must be integer quantized momenta conjugate to the $2\pi$-periodic holonomies $\int_{\Sigma_i}B$. We also expand $H = m^i \t \Omega_i$ where $m^i$ are integer quantized magnetic charges and $\t \Omega_i$ are dual three-forms to $\Omega^i$ as defined in Appendix $A$. By taking
\bea
\lambda &=& \frac{1}{4\pi}
\eea
we get the zero mode part of the Hamiltonian as
\bea
H_{zero} &=& \sqrt{-\beta} 2\pi \G_{ij} \(n^i n^j + \frac{1}{4} m^i m^j\) - 2\pi \L_{ij} n^i m^j
\eea
where $\G_{ij}$ denotes the inverse of $\G^{ij}$ and $\L_{ij}$ is defined in Appedix $A$. The zero mode partition function is defined as
\bea
\Z_{zero} &=& \sum_{n^i,m^i} e^{-2\pi i H_{zero}}
\eea
By identifying the period matrix as
\bea
\tau_{ij} &=& 2\pi \(-\L_{ij} + \sqrt{-\beta} \G_{ij}\)
\eea
the zero mode partition function can be written as\footnote{Complex conjugation works as usual only in Euclidean signature where $\beta$ is real and positive. But we can extend analytically to arbitrary complex $\beta$ while always prescribing the conjugation rule $\overline{\sqrt{-\beta}} = -\sqrt{-\beta}$. Using this prescription we can use theta function formalism also in Minkowski signature which is the natural signature for Hamiltonian quantization.}
\bea
\Z_{zero} &=& \sum_{a^i,b_i} \Theta\[{}^a_b\](-\tau) \overline{\Theta\[{}^a_b\](-\tau)}
\eea
Here the Jacobi theta functions are given by
\bea
\Theta\[{}^a_b\](\tau) &=& \sum_{n^i\in \mb{Z}} \exp\( \pi i (n^i+a^i) \tau_{ij} (n^j+a^j) + 2\pi i n^i b_i \)
\eea
and the characteristics $a^i$ and $b_i$ are running over $0, \frac{1}{2}$.

So far we have computed the zero mode part of the partition function of a non-selfdual three-form $H$, and we have seen that it is holomorphically factorizable in a certain sense, as a finite sum of products of chiral and antichiral parts. On the Chern-Simons theory side, this corresponds to the inner product of wave functions. The wave function that corresponds to the partition function of selfdual three-form is now given by
\bea
\Z &=& \Theta\[^a_b\](-\tau|0)
\eea
for some certain characteristics.

This computation shows that we shall choose the corresponding Chern-Simons level to be $k=1$ but it might be interesting to ask whether other values of $k$ can be implemented in M5 brane theory as well.

\subsection{An alternative treatment of selfdual tensor gauge field}\label{Perry-Schwarz}
For the tensor gauge field we may use a selfdual Lagrangian. Holomorphic factorization works nicely for the zero modes if we define the period matrix on the space of harmonic three-forms. For the oscillator modes this method requires an extension of the period matrix to include the oscillator modes. However this extended period matrix would be an infinite-dimensional matrix which could lead to additional subtleties in the holomorpic factorization. The alternative route is to work directly with a selfdual Lagrangian from which we can compute the contribution of the the selfdual oscillator modes using conventional quantization methods, that is Hamiltonian quantization or path integral quantization.

Let us first return to the nonselfdual Lagrangian. For the zero modes we have two set of integers, $m^i$ and $n^i$. Using notations introduced in Appendix $A$, these integers were defined as
\bea
H_{mnp} &=& m^i \t \Omega_{i,mnp}\cr
e^{mn} &=& \frac{n^i}{2} \Omega_{i}^{mn}
\eea
Let us now decompose the three-form into selfdual and antiselfdual parts,
\bea
H &=& H^+ + H^-
\eea
where
\bea
H^{\pm,tmn} &=& \pm \frac{\sqrt{-g}}{6} \epsilon^{tmnpqr} H^{\pm}_{pqr}
\eea
We accordingly decompose the two sets of integers as
\bea
m^i &=& m^{i+} + m^{i-}\cr
2n^i &=& m^{i+} - m^{i-}
\eea
whose solutions are
\bea
m^{i\pm} &=& n^i \pm \frac{m^i}{2}\cr
\eea
We can express this as saying that
\bea
m^{i+} &=& p^i + a^i
\eea
where $p^i \in \mb{Z}$ and each $a^i$ is either $0$ or $\frac{1}{2}$.

We now turn to the selfdual Lagrangian. We define a six-dimensional vielbein $e^A = e^A_M dx^M$ which we decompose into $e^A = (e^t,e^{\alpha})$ where $\alpha =1,\cdots,5$. For the three-form field strength we have
\bea
H_{MNP} &=& e^A_M e^B_N e^C_P H_{ABC}
\eea
where we define
\bea
\partial_A &=& e_A^M \partial_M
\eea
and
\bea
H_{ABC} &=& 3\partial_{[A} B_{BC]}
\eea
To see this one notices that the ordinary derivative can be replaced covariant derivatives in the three-form, and the vielbein is covariantly constant. Then we consider the following six-dimensional Lagrangian ($\lambda = \frac{1}{4\pi}$)
\bea
L &=& -\frac{\lambda}{12} \int d^5 x \sqrt{G} \epsilon^{\alpha\beta\gamma\delta\epsilon} H_{\alpha\beta\gamma} \partial_t B_{\delta\epsilon} - \frac{\lambda\sqrt{-\beta}}{6} \int d^5 x \sqrt{G}H^{\alpha\beta\gamma} H_{\alpha\beta\gamma}
\eea
As we showed above, the Dirac charge quantization for the selfdual field strength is given by
\bea
\frac{1}{2\pi}\int_{c_i} H &=&  {\mb{Z}} + a^i
\eea
over spatial three-cycles $c_i$. Moreover, the selfdual holonomies
\bea
X_i(t) &=& \exp^{i\int_{\Sigma_i} B^+}
\eea
are periodic or antiperiodic according to
\bea
X_i(t+2\pi) &=& (-1)^{b_i} X_i(t)
\eea
The $b_i$ correspond to the characteristics $b_i$ in the theta function. In two-dimensions the holonomies correspond to fermions via bosonization.

\section{The D4 brane partition function}
If we choose the metric on the M5 brane worldvolume on the form
\bea
ds^2_{M5} &=& \t G_{\mu\nu} dx^{\mu} dx^{\nu} + R^2 (dx^5 + v_{\mu} dx^{\mu})^2
\eea
then the corresponding D4 brane action will be given by
\bea
S &=& S_{YM} + S_{\phi'} + S_{\psi'}
\eea
where
\bea
S_{YM} &=& -\frac{1}{4 g^2} \int d^5 x \sqrt{-\t G} F^{\mu\nu} F_{\mu\nu} + \frac{1}{32\pi^2} \int d^5 x \t G \epsilon^{\mu\nu\kappa\tau\sigma} F_{\mu\nu} F_{\kappa\tau} v_{\sigma}\cr
S_{\phi'} &=& -\frac{1}{2g^2} \int d^5 x \sqrt{-\t G} \partial^{\mu} \phi'^A \partial_{\mu} \phi'^A\cr
S_{\psi'} &=& \frac{i}{g^2} \int d^5 x \sqrt{-\t G} \bar{\psi}' \Gamma^{\mu} \partial_{\mu} \psi'
\eea
and the coupling constant is
\bea
g^2 &=& 4\pi^2 R
\eea

\subsection{Scalar field}
The scalar field action is given by
\bea
S &=& -\frac{1}{2g^2} \int d^5 x \sqrt{-\t G} \t G^{\mu\nu} \partial_{\mu} \phi' \partial_{\nu} \phi'
\eea
where
\bea
g^2 &=& 4\pi^2 R
\eea
This action is obtained by dimensional reduction of (\ref{scalar}) and by defining
\bea
\phi' &=& 2\pi R \phi
\eea
We introduce a regulator $r'$ such that
\bea
\phi' &\sim & \phi' + 2\pi r'
\eea
but of course this regulator is different from the M5 brane regulator $r$. These are related as
\bea
r' &=& 2\pi R r
\eea
The quickest way to compute the zero mode contribution to the partition function is by lending the result from the M5 brane. To this end we rescale the scalar field so that the action gets the factor $-\lambda$ in front in place of $\frac{1}{2g^2}$. We then find the periodicity gets modified to
\bea
\phi' &\sim & \phi' + 2\pi r''
\eea
with 
\bea
r'' &=& \frac{1}{\sqrt{2\pi R}} 2\pi r'
\eea
Then we can borrow the M5 brane result and we get the zero mode part of the Hamiltonian as
\bea
H_{D4} &=& \frac{\lambda\sqrt{-\beta}}{\Vol'} \frac{n^2}{r''^2}
\eea
where we use $\Vol' = \int d^4 x \sqrt{g}$ in place of $Vol = \int d^5 x \sqrt{G}$. These volume factors are related by
\bea
\Vol &=& 2\pi R \Vol'
\eea
To compare with the M5 brane Hamiltonian, we substitute $r''=\sqrt{2\pi R} r$. Then
\bea
H_{D4} &=& \frac{\lambda\sqrt{-\beta}}{\Vol'2\pi R}\frac{n^2}{r^2}
\eea
which now agrees with the zero mode part of the M5 brane Hamiltonian. Consequently also $\Z_{zero,\phi'}$ agrees with $\Z_{zero,\phi}$ of the M5 brane.

\subsection{Gauge field}
The momentum conjugate to $A_i$ is
\bea
E^i &=& -\frac{1}{4\pi^2 R} \sqrt{-\t G} F^{ti} + \frac{1}{8\pi^2} \t G\epsilon^{tijkl} F_{jk} v_l
\eea
We choose temporal gauge
\bea
A_t &=& 0
\eea
We first show that we can express the conjugate momentum in a six-dimensional form as
\bea
E^{i} &=& -\frac{1}{2\pi}\sqrt{-g} H^{5ti}_{6D}
\eea
To see this, we first note that
\bea
H^{5\mu\nu}_{6D} &=& \frac{1}{R^2} \t G^{\mu\kappa} \t G^{\nu\lambda} H_{5\kappa\lambda} - v_{\lambda} H^{\mu\nu\lambda}_{6D}
\eea
Then using selfduality, we have
\bea
H^{5\mu\nu}_{6D} &=& \frac{1}{R^2} \t G^{\mu\kappa} \t G^{\nu\lambda} H_{5\kappa\lambda} - \frac{1}{2} \sqrt{-g} \epsilon^{\mu\nu\lambda\kappa\tau 5} v_{\lambda} H_{\kappa\tau 5}
\eea
and then
\bea
R \sqrt{-\t G} H^{ti5}_{6D} &=& \frac{1}{2\pi} \(\frac{1}{R} \sqrt{-\t G} F^{ti} + \frac{1}{2} \t G \epsilon^{tijkl} v_j F_{kl}\)
\eea
where we define
\bea
H_{\mu\nu 5} &=& \frac{1}{2\pi} F_{\mu\nu}
\eea
and the identity follows.

We next comment on the integer quantization of the momentum which follows by the fact that it is conjugate to $2\pi$-periodic holonomies. For
YM we then have
\bea
\[\int_0^{2\pi} dx^1 A_1,E^1\] &=& 2\pi i
\eea
and for the six-dimensional theory we have
\bea
\[\int_0^{2\pi} dx^1 \int_0^{2\pi} dx^5 B_{15},2\pi E^{15}\] &=& 2\pi^2 i
\eea
where we write $2\pi E^{15}$ due to integration over $x^5$ as necessary to get $\delta^4$ quantities on the rhs in both cases. From these relations together with $2\pi$-periodicity of holonomies, we conclude that
\bea
E^1 &=& 2\pi n^1\cr
2\pi E^{15} &=& 2\pi^2 n^{15}
\eea
where $n^1$ and $n^{15}$ are integer quantized. By identifying these integers as $n^i = n^{i5}$ for $i=1,2,3,4$, we conclude that
\bea
2\pi E^{i5} &=& \pi E^i
\eea
and thus we should expect the conjugate momentum of our `would be' six-dimensional theory, should be given by
\bea
E^{mn} &=& -\frac{1}{4\pi}\sqrt{-g} H^{tmn}_{6D}
\eea
or in other words,
\bea
E^{mn} &=& -\lambda\sqrt{-g} H^{tmn}_{6D}
\eea

The Hamiltonian we derive from the 5D YM is given by
\bea
\H_{YM} &=& E^i F_{ti} - \L_{YM}\cr
\eea
The first term is
\bea
E^i F_{ti} &=& -\sqrt{-g} H_{6D}^{ti5} H_{ti5}
\eea
For the second term we first notice that
\bea
-\frac{1}{4}\sqrt{-g} H_{6D}^{5\mu\nu} H_{5\mu\nu} &=& \L_{YM}
\eea
and furthermore
\bea
-\frac{1}{4}\sqrt{-g} H_{6D}^{5\mu\nu} H_{5\mu\nu} &=& -\sqrt{-g} \(\frac{1}{4}H_{6D}^{5ij} H_{5ij} + \frac{1}{2}H_{6D}^{5ti} H_{5ti}\)
\eea
Thus
\bea
\H_{YM} &=& \sqrt{-g} \(\frac{1}{4}H_{6D}^{5ij} H_{5ij} - \frac{1}{2}H_{6D}^{5ti} H_{5ti}\)
\eea

The six-dimensional Hamiltonian is given by
\bea
H_{M5} &=& \frac{\lambda R}{6} \int d^5 x \sqrt{-\t G} H_{6D}^{mnp} H_{mnp}
\eea
but by using the selfduality constraint this Hamiltonian can also be expressed as
\bea
H_{M5} &=& -\frac{\lambda R}{2} \int d^5 x \sqrt{-\t G} H_{6D}^{tnp} H_{tnp}\cr
&=& R \int d^4 x \sqrt{-\t G} \(-\frac{1}{2} H^{tij}_{6D} H_{tij} + \frac{1}{4} H^{5ij}_{6D} H_{5ij}\)
\eea
To get to the second line we have assumed that the fields are independent of the coordinate on the fiber, so that integration along the fiber just gives a factor of $2\pi$. We also used selfduality again, now in the form of $H^{tij}_{6D} H_{tij} = -H^{5ij}_{6D} H_{5ij}$. We now have
\bea
H_{M5} &=& H_{YM}
\eea
under the assumption of fields being independent on the fiber coordinate. In particular then, the zero mode partition functions are identical.

\section{Oscillators}

\subsection{Selfdual tensor field}
In the spirit that M5 brane theory is a higher dimensional analog of selfdual scalar field in two dimensions, we begin by computing the oscillator partition function for a two-dimensional selfdual scalar field in a way that is easily generalized to a selfdual two-form gauge potential in six dimensions.

\subsubsection{Two-torus}
Let us consider a real scalar field on $T^2$ with metric
\bea
ds^2 = \beta dt^2 + G_{xx} (dx - V^x dt)^2 = G_{xx} (dx + \tau dt) (dx + \bar{\tau} dt)
\eea
where
\bea
\tau = -V^x + \frac{\sqrt{-\beta}}{\sqrt{G_{xx}}}\cr
\bar{\tau} = - V^x - \frac{\sqrt{-\beta}}{\sqrt{G_{xx}}}
\eea
The action is
\bea
S &=& -\frac{1}{2} \int dt dx \sqrt{-g} g^{MN}\partial_M \phi \partial_N \phi
\eea
As the orthonormal basis on the base-manifold whose coordinate is $x \sim x + 2\pi$ we take $\varphi_n = \frac{1}{2\pi} e^{inx}$ for $n\geq 0$. The complex conjugates are $\varphi^n = \varphi_{-n}$. The quadratic equation of motion reads (for $n\geq 0$)
\bea
\(\frac{1}{\beta} \(\partial_t + V^x \partial_x\)^2 + G^{xx} \partial_x^2\) e^{iEt + inx} &=& 0
\eea
and it has solutions
\bea
E^+_n &=& \tau n\cr
E^-_n &=& \bar{\tau} n
\eea
In Minkowski signature $E^+_n \geq 0$ and $E^-_n \leq 0$. The Hamiltonian is
\bea
H &=& \frac{1}{2} \sum_{n \in \mb{Z}_+} n \[\tau \(a^{\dag}_n a_{n} + \frac{1}{2}\) - \bar{\tau} \(b^{\dag}_n b_{n} + \frac{1}{2}\)\]
\eea
where
\bea
[a_m,a^{\dag}_{n}] &=& \delta_{m,n}\cr
[b_m,b^{\dag}_{n}] &=& \delta_{m,n}
\eea
The partition function becomes
\bea
Z_{osc} &=& \frac{1}{\eta(\tau)\overline{\eta(\tau)}}
\eea
where
\bea
\eta(\tau) &=& e^{-\frac{i\pi \tau}{12}}\prod_{n=1}^{\infty} \(1 - e^{2\pi i \tau n}\)
\eea
is the Dedekind eta function.

We note that the partition function can be expressed as a determinant
\bea
Z_{osc} &=& \frac{2\pi \sqrt{\beta}}{\sqrt{\det' \triangle}}
\eea
where in this determinant we shall put $\partial_t = im$ where $m\in \mb{Z}$ due to our convention where the time coordinate interval over which we compute the partition function is $[0,2\pi]$. The physical time interval is $2\pi \sqrt{|\beta|}$.

\subsubsection{Six-manifold}
We now proceed to generalize this to six dimensions. If we impose temporal and Lorentz gauge
\bea
i_t B &=& 0\cr
d^{\dag} B &=& 0
\eea
with respect to six-dimensional metric, then the Maxwell equation of motion becomes
\bea
\triangle_6 B &=& 0
\eea
where $\triangle_6$ denotes the Laplacian on $M_6$. The energies are obtained by solving the equation
\bea
\triangle_6 \(e^{iE_{\alpha} t}\varphi_{\alpha}\) &=& 0
\eea
where $\varphi_{\alpha}$ denote an orthonormal basis of two-forms on $M_5$. Such a basis can be conveniently chosen such that
\bea
\triangle_5 \varphi_{\alpha} &=& \lambda_{\alpha} \varphi_{\alpha}\cr
(\varphi_{\alpha},\varphi_{\beta}) &=& \delta_{\alpha}^{\beta}
\eea
where the inner product is given by $(\varphi_{\alpha},\varphi_{\beta}) = \int_{M_5} \varphi_{\alpha} \wedge *_5 \bar{\varphi}_{\beta}$ and $\lambda_{\alpha}$ denote eigenvalues and bar denotes complex conjugation.

Obviously the eigenvalues $\lambda_{\alpha}$ may be degenerate. What is less obvious is that the partition function can be complex-valued. To see this, let us assume a Cartesian product $M_6 = T^2 \times M_4$. In this case we get
\bea
B &=& \phi(t,x) \varphi(x^i)
\eea
and we wish to solve the equation of motion
\bea
(\triangle_2 \phi) \varphi + \phi \triangle_4 \varphi &=& 0
\eea
Let us assume that we solve this equation by requiring
\bea
\triangle_4 \varphi &=& 0
\eea
Then $d\varphi = 0$ and we get
\bea
dB &=& d\phi \wedge \varphi
\eea
Requiring this to be selfdual means that either both $\varphi$ and $d\phi$ are selfdual or both antiselfdual. This shows that the partition function will now contain a factor
\bea
\Z_{osc}(\tau,g_{ij}) &=& \frac{1}{{\eta(\tau)}^{b_+}{\overline{\eta(\tau)}}^{b_-}}
\eea
where $b^{\pm}$ denote the number of selfdual and antiselfdual harmonics on $M_4$ respectively, and $\eta(\tau)$ is the contribution from one selfdual scalar field on $T^2$. Only if $b^+ - b^- = 0$ do we have a real-valued oscillator partition function. The oscillator mode contribution for a selfdual tensor gauge field is given by the holomorphic factor $\frac{1}{{\eta(\tau)}^{b_+}}$. We do not know how this factorization extends beyond the zero modes on $M_4$ in this case. One way to compute the oscillator mode contribution to the partition function of a selfdual tensor gauge field could be to directly work with a selfdual action for example the one we wrote down in section \ref{Perry-Schwarz}.

\subsubsection{Six-torus}
Let us now instead turn to the case when $M_6$ is flat $T^6$. Since this is $T^2 \times T^4$ and $b^+(T^4) = b^-(T^4) = 3$ the oscillator partition function will be real. This means that to compute the oscillator mode partition function for a selfdual tensor gauge field on $T^6$ we can simply take the square root of the oscillator mode partition function of a non-selfdual tensor gauge field on $T^6$. The Laplacian is
\bea
\triangle_6 &=& -g^{MN} \partial_M \partial_N
\eea
and this gets split into
\bea
\triangle_6 &=& -\frac{1}{\beta} \partial_t^2 - \frac{2}{\beta} V^m \partial_t \partial_m + \triangle_5
\eea
Let us define
\bea
f^{mn}_{\p} &=& \frac{1}{(2\pi)^{\frac{5}{2}}} e^{i p_m x^m} dx^m \wedge dx^n
\eea
which satisfy
\bea
\triangle_5 f^{mn}_{\p} &=& \lambda_{\p} f^{mn}_{\p}\cr
\(f^{mn}_{\p},f^{pq}_{\q}\) &=& \sqrt{G} G^{mn,pq} \delta_{\p,\q}
\eea
where
\bea
\lambda_{\p} &=& p^2\cr
p &:=& \sqrt{G^{mn} p_m p_n}
\eea
We wish the consider a basis $f_{\vec{p}}^{mn}$ with the property that $(f_{\vec{p}}^{mn})^*$ are linearly independent of $f_{\vec{p}}^{mn}$. This can be achieved by restricting ourselves to a half-space $\vec{p} \in \mb{Z}_+^5$. This half-space can be selected by first picking a co-dimension one hyperplane in $\mb{R}^5$ which only intersects the origin but no other lattice points in $\mb{Z}^5$. Such a hyperplane may be defined by one normal vector which we may take for example as $\vec{n} = (1,\sqrt{2},\sqrt{3},\sqrt{5},\sqrt{7},\sqrt{11})$. Then we may define the half-space as $\mb{Z}_+^5 = \{\vec{p} \in \mb{Z}^5 | \vec{n}\cdot \vec{p} >0\}$. The energies are
\bea
E^{\pm}_{\p} &=& -V^m p_m \pm \sqrt{-\beta} p
\eea
Let us define a projector
\bea
(\Pi_{\p})_m^n &=& \frac{1}{2} \(\delta_m^n - \hat{p}_m \hat{p}^n\)
\eea
Then
\bea
\t{f}^{mn}_{\p} &=& (\Pi_{\p})^{mn}_{pq} f^{pq}_{\p}
\eea
satisfy Lorentz gauge $p_m f^{mn}_{\p} = 0$ (we drop the tilde for notation convenience). Moreover
\bea
\(f^{mn}_{\p},f^{pq}_{\q}\) &=& \sqrt{G} \Pi^{mn,pq} \delta_{\p,\q}
\eea
We also define selfduality projectors
\bea
(S_{\p})^{mn}_{pq} &=& \frac{1}{2} \(\Pi^{mn}_{pq} + \frac{1}{2}\sqrt{G}\epsilon^{mn}{}_{pqr}\hat{p}^r\)
\eea
and define
\bea
\t{f}^{mn}_{\p} &=& (S_{\p})^{mn}_{pq} f^{pq}_{\p}
\eea
which obey (again dropping the tilde)
\bea
\(f^{mn}_{\p},f^{pq}_{\q}\) &=& \sqrt{G} S^{mn,pq} \delta_{\p,\q}
\eea
We expand
\bea
B(x) &=& \sum_{\p \in {\mb{Z}}^5_+} \(a_{mn,\p} e^{iE^+_{\p}t} + b_{mn,\p} e^{iE^-_{\p}t}\) f^{mn}_{\p}(\x) + c.c.
\eea
and this will now satisfy the Maxwell equation of motion, be in Lorentz gauge and give a selfdual field strength. We see that there are three independent polarizations $a_{mn,\p}$ for a given momentum $\p$. Let us denote these independent ones as $a^i_{\p}$ for $i=1,2,3$.

It is a general result that the Hamiltonian of a quadratic Lagrangian is given by
\bea
H &=& \sum_{\p\in \mb{Z}_+^5,i} \[E^+_{\p} \(a^{i\dag}_{\p} a^i_{\p} +\frac{1}{2}\) - E^-_{\p} \(b^{i\dag}_{\p} b^i_{\p} +\frac{1}{2}\)\]
\eea
where
\bea
[a^i_{\p},a^{j\dag}_{\q}] &=& \delta^{ij}\delta_{\p,\q}\cr
[b^i_{\p},b^{j\dag}_{\q}] &=& \delta^{ij}\delta_{\p,\q}
\eea
Since the Hamiltonian shall be non-negative in Minkowski signature, we shall take $E^+$ to be non-negative, and $E^-$ to be non-positive solutions to the quadratic equation of motion. The oscillator partition function now becomes
\ben
\Z_{osc} &=& \frac{1}{\eta(g^{MN})^3\overline{\eta(g^{MN})^3}}\label{osc_tensor}
\een
where we define a generalized Dedekind eta function and its conjugate as
\bea
\eta(g^{MN}) &=& e^{-i\pi \sum_{\p \in {\mb{Z}}^5_+} E^+_{\p}} \prod_{\p\in {\mb{Z}}^5_+} \(1 - e^{2\pi i E_{\p}^+}\)\cr
\overline{\eta(g^{MN})} &=& e^{i\pi \sum_{\p \in {\mb{Z}}^5_+} E^-_{\p}} \prod_{\p \in {\mb{Z}}^5_+} \(1-e^{-2\pi i E^-_{\p}}\)
\eea
If we pick $T^2$ with modular parameter $\tau$ embedded in $T^6$ then $
Z_{osc}$ contains as a factor $\frac{1}{\eta(\tau)^3 \overline{\eta(\tau)}^3}$. This is an important observation to show modular invariance of the full partition function on $T^6$.

\subsection{Scalar fields}
For the five scalar fields we get the contribution
\bea
\frac{1}{\eta(g^{MN})^5\overline{\eta(g^{MN})^5}}\label{osc_scalar}
\eea
where $E^{\pm}$ are computed from $\triangle_6$ acting on zero forms $e^{iEt} \phi(x^m)$. For a flat $T^6$ we get that same eta functions as those for the selfdual tensor field since for flat space we have the same energies.

\subsection{Fermions}
In order to preserve supersymmetry on $T^6$ we shall impose periodic boundary conditions on the fermions on all the $6$ one-cycles of $T^6$. For the spatial directions this amounts to integer momenta. To get periodic fermions in the time direction we must insert $(-1)^F$. Thus the fermionic partition function is
\bea
\tr((-1)^F e^{2\pi i H}) &=& e^{-8i\pi \sum_{\p \in {\mb{Z}}^5} \(E_{\vec{p}}^+ - E_{\vec{p}}^-\)} \prod_{\p\in {\mb{Z}}^5_+ +\, \vec{0}} \(1 - e^{2\pi i E_{\p}^+}\)^8 \(1 - e^{-2\pi i E_{\p}^-}\)^8
\eea
Now for this periodic boundary conditions there is a zero mode $\p=0$ which kills the whole partition function so that
\bea
\Z &=& 0
\eea
Obviously this partition function is modular invariant. We may also relax period boundary condition in time direction while preserving supersymmetry since supersymmetry variations are at a fixed time and do not see such a boundary condition in time. In that case we get the partition function as
\bea
\tr(e^{2\pi i H}) &=& e^{-8i\pi \sum_{\p \in {\mb{Z}}^5} \(E_{\vec{p}}^+ - E_{\vec{p}}^-\)}\prod_{\p\in {\mb{Z}}^5_+ +\, \vec{0}} \(1 + e^{2\pi i E_{\p}^+}\)^8 \(1 + e^{-2\pi i E_{\p}^-}\)^8
\eea
but now we can only expect this to be modular invariant under modular transformations that act on $T^5 \subset T^6$ and leave the time direction unaffected. So we shall expect this partition function to be only $SL(5,\mb{Z})$ invariant.

\section{Modular invariance of M5 brane on $T^2 \times M_4$}\label{modular invariance}
So far we have been studying M5 brane on circle fibration over a five-manifold. If we continue the program of circle fibrations, the next step will be to consider a two-torus fibered over some four-manifold. Since the modular group of a two-torus is $SL(2,\mb{Z})$ we can ask whether the M5 brane partition function is $SL(2,\mb{Z})$ invariant. Such a strategy was used in \cite{Dolan:1998qk} for the case of a flat $T^6$ to prove full $SL(6,\mb{Z})$ modular invariance on $T^6$ by first proving $SL(2,\mb{Z})$ modular invariance on a $T^2$ embedded in $T^6$. But we may generalize. The most general geometry for which $SL(2,\mb{Z})$ modular invariance can be studied is where we have two commuting Killing vectors corresponding to a two-torus $T^2$ in the six-manifold $M_6$. If we take one circle direction of that $T^2$ to be associated with time, the natural choice of metric will be
\bea
ds^2 &=& \beta dt^2 + G_{mn} (dx^m - V^m dt)(dx^n - V^n dt)
\eea
on which we make a subsequent decomposition of the same form,
\bea
G_{mn} dy^m dy^n &=& \gamma (dy^5)^2 + g_{ij} (dy^i - U^i dy^5)(dy^j - V^j dy^5)
\eea
Thus, by substituting $dy^m = dx^m - V^m dt$, the full M5 brane metric is
\bea
ds^2 &=& \beta dt^2 + \gamma (dx - V^5 dt)^2 + g_{ij} (dx^i - V^i dt - U^i dx)(dx^j - V^j dt - U^i dx)
\eea
The first two terms can be combined into the standard metric on a two-torus
\bea
\gamma(dx + \tau dt)(dx + \bar{\tau} dt)
\eea
if we define
\bea
\tau &=& - V^5 +  \frac{\sqrt{-\beta}}{\sqrt{\gamma}}
\eea
The metric is invariant under simultaneous exchange of coordinates on $T^2$,
\bea
t' &=& x\cr
x' &=& -t
\eea
and the following change of parameters in the metric,
\bea
V'^i &=& U^i\cr
U'^i &=& -V^i\cr
\tau' &=& - \frac{1}{\tau}\cr
\bar{\tau}' &=& - \frac{1}{\bar{\tau}}\cr
\gamma' &=& \tau \bar{\tau} \gamma
\eea
We will postpone a study of modular invariance on generic such fiber-bundles to future work. Here we will assume the six-manifold $M_6$ is on the form of a Cartesian product
\bea
M_6 &=& T^2 \times M_4
\eea
This corresponds to putting $V^i = U^i = 0$ above. The metric on $M_6$ is then
\bea
ds^2_{M_6} &=& ds^2_{M_2} + ds^2_{M_4}
\eea
where
\bea
ds^2_{M_2} &=& \gamma \(dx + \tau dt\) \(dx + \bar{\tau} dt\)\cr
ds^2_{M_4} &=& g_{ij}dx^i dx^j
\eea
The Laplace operator becomes
\bea
\triangle_6 &=& \triangle_2 + \triangle_6
\eea

\subsection{The zero mode part}
In the Appendix \ref{period2} we obtain the period matrix for $T^2 \times M_4$ for the case that dim$H^1(M_4)=0$. This period matrix has been previously obtained in \cite{Henningson:2000gu}. We have the following results for the period matrix and its inverse,
\bea
\tau_{IJ} &=& \tau_1 (Q^{-1})_{IJ} + \tau_2 G_{IJ}\cr
(\tau^{-1})^{IJ} &=& \t{\tau}_1 Q^{IJ} + \t \tau_2 Q^{IK}Q^{JL}G_{KL}
\eea
Here $\tau$ is the period matrix on $T^2$ and $Q^{IJ}$ and $G^{IJ}$ is the intersection form and the metric associated with $M_4$ as defined in the Appendix \ref{period2}. We define\footnote{We remind that our notation is $\tau = -\tau_1 - \tau_2$ and $\bar{\tau} = -\tau_1 + \tau_2$ in both Minkowski and Euclidean signatures. We then define $\t \tau = -\frac{1}{\tau} = -\frac{\bar{\tau}}{\tau \bar{\tau}}$.}
\bea
\t \tau_1 &=& -\frac{\tau_1}{\tau_1^2 - \tau^2_2}\cr
\t \tau_2 &=& -\frac{\tau_2}{\tau_1^2 - \tau^2_2}
\eea
The partition function is\footnote{For simplicity we consider only zero characteristics. It is also true that this choice is the most modular invariant choice \cite{Bonelli:2001pu}, \cite{Gustavsson:2000kr}}
\bea
\Z_{zero} &=& \Theta\[{}^0_0\](\tau_{IJ})
\eea
Two ingredients are involved to see the modular property of this partition function under $S$-transformation. The first ingredient is Poisson resummation,
\bea
\sum_{n^I \in \mb{Z}} e^{i\pi n^I \tau_{IJ} n^J} &=& \frac{1}{\sqrt{\det \tau_{IJ}}} \sum_{m_I \in \mb{Z}} e^{i\pi m_I (\tau^{-1}) m_J}
\eea
The second ingredient is to make use the intersection form (whose entries are integers) to define dual integers
\bea
m^I &=& Q^{IJ} m_J
\eea
which results in
\bea
m_I (\tau^{-1}) m_J &=& m^I \t \tau_{IJ} m^J
\eea
where we define
\bea
\t \tau_{IJ} &=& \t \tau_1 (Q^{-1})_{IJ} + \t \tau_2 G_{IJ}
\eea
To compute the determinant we define
\bea
m &=& m_I \Omega^I
\eea
and
\bea
m^{\pm} &=& m \pm * m
\eea
we then consider the following equality (noting that $** = 1$ and defining the inner product as usual).
Then
\bea
m_I Z^{IJ} m_J &=& \frac{1}{4} \(\tau (m^+,m^+) + \bar{\tau} (m^-,m^-)\)
\eea
We can compute the determinant as a product of exponents. These exponents factorize into selfdual parts, so we can compute the determinant of each factor. One such factor is
\bea
\frac{1}{\sqrt{\det Z^{IJ}}} &=& \prod_{m^+} \int dm^+ \exp \frac{1}{4}\tau (m^+,m^+)
\cr&=& \(\frac{1}{\sqrt{\tau}}\)^{b_+}
\eea
The product runs over the selfdual harmonic two-forms, and $b_+$ is the number of such two-forms.

No contribution relevant for S-duality comes from harmonic one-forms on $M_4$ because these do not give a dependence on $\tau$. They involve various combinations that contain $dt\wedge dx$, but no single $dx$ nor single $dt$, so there is no $\tau$ dependence.

We find that the zero mode partition function transforms like
\bea
\Z_{zero}(-\frac{1}{\tau},g_{ij}) &=& {\sqrt{\tau}}^{b_+}{\sqrt{\bar{\tau}}}^{b_-}\Z_{zero}(\tau,g_{ij})\cr
\Z_{zero}(\tau+1,g_{ij}) &=& \Z_{zero}(\tau,g_{ij})
\eea
A corresponding result was obtained in \cite{Henningson:2000gu} for non-Abelian gauge group. However in this reference it seems the second ingredient of using the intersection form $Q^{IJ}$ to define dual integer numbers is missing.

\subsection{Oscillator mode part}
Oscillators contribute with a generalized eta function. Zero modes have already been analyzed, and the determinant of the six-dimensional Laplacian acting on two-forms, equals the generalized eta function up to a zero mode factor.

We have learned how to extract $\Z_{osc}$ from the determinant. Let us now consider the determinant that is relevant to this situation
\bea
\det \triangle_6 &=& \det (\triangle_2 + \triangle_4)
\eea
We will ignore the issue of gauge fixing and ghost determinants. Instead we will map this determinant directly to $\Z_{osc}$. Only massless modes from two-dimensional point of view can contribute an $SL(2,\mb{Z})$ anomaly. So it is enough to consider zero modes of $\triangle_4$. These are the harmonic two-forms on $M_4$. Thus the determinant we will be interested in reads
\bea
(\det \triangle_2)^{b_2}
\eea
From this result we can immediately conclude that the $\tau$ dependence of the oscillator partition function is
\bea
\Z_{osc}(\tau,g_{ij}) &=& \frac{1}{{\eta(\tau)}^{b_+}{\overline{\eta(\tau)}}^{b_-}}
\eea
and it transforms as
\bea
\Z_{osc}(-\frac{1}{\tau},g_{ij}) &=& \(\frac{1}{\sqrt{\tau}}\)^{b_+}\(\frac{1}{\sqrt{\bar{\tau}}}\)^{b_-}\Z_{osc}(\tau,g_{ij})\cr
\Z_{osc}(\tau+1,g_{ij}) &=& e^{-\frac{i\pi}{12}(b_+-b_-)} \Z_{osc}(\tau,g_{ij})
\eea
The total partition function is thus invariant under $S$-transformation $\tau \rightarrow -\frac{1}{\tau}$ and transforms at most by a phase under $T$-transformation $\tau\rightarrow \tau + 1$. The partition function is modular invariant whenever $b_+ = b_-$. This shows that the partition function is modular invariant on $T^6$ but not on $T^2 \times TN$ where $b_+ \neq b_-$. (On $TN$ we have $b_+ = 1$ and $b_- = 0$).

For the five scalar fields we also find modular invariance on $T^6$, despite the oscillator mode contribution by itself is not modular invariant. But we do have a zero mode contribution also for the non-compact scalar fields, as we showed in section \ref{scalarzeromode}. For the case that $W = T^2 \times M_4$ (whereof $T^6$ is a special case) we find that we can express the zero mode factor as
\bea
\Z_{zero} &=& \(\frac{\gamma}{-\beta}\)^{\frac{1}{4}} \cr
&=& \frac{1}{\sqrt{\tau_2}}
\eea
It transforms under $S$-transformation into
\bea
\Z_{zero}(-\frac{1}{\tau},g_{ij}) &=& |\tau| \Z_{zero}(\tau)
\eea
For a single scalar field the oscillator mode partition function transforms as
\bea
\Z_{osc}(-\frac{1}{\tau},g_{ij}) &=& \frac{1}{|\tau|} \Z_{osc}(\tau)
\eea
showing that $\Z_{zero} \Z_{osc}$ is invariant under $S$-transformation. The same holds true for $T$-transformation on $T^6$.

For the fermions on $T^6$ and if we assume antiperiodic boundary conditions in all directions, we find a zero mode that kills the whole partition function. Obviously $0$ is $SL(6,\mb{Z})$ modular invariant. If we choose periodic boundary conditions on the circle in $T^6$ that we associate with time, then we shall only expect to find $SL(5,\mb{Z})$ invariance. And indeed this amount of symmetry is manifest in our expression for the oscillator mode partition function, as was also noted in \cite{Dolan:1998qk} in the context of tensor gauge field on $T^6$.

Thus the M5 brane partition function with zero characteristics on a flat $T^6$ is $SL(6,\mb{Z})$ modular invariant, and on $T^2 \times M_4$ it is $SL(2,\mb{Z})$ modular invariant up to a phase. Under dimensional reduction to four dimensions we find a partition function which is a modular form \cite{Witten:1995gf}. It is natural that we get  different modular properties of the partition functions in four and six dimensions since we truncate away all the KK modes in four dimensions. Nevertheless we do find a modular property in four dimensions. This experience could motivate us to ask if some modular property could also be found in five dimensions.

The oscillator parts of both D4 and M5 are on the form of generalized Dedekind eta functions. Let us consider a flat six-torus. Then the frequencies $E_{n_m}$ that appear in the generalized Dedekind eta function are the positive roots to the equation of motion
\bea
g^{MN} n_M n_N &=& 0
\eea
where we solve this equation for $n_t = E_{n_m}$ and pick the positive solutions (in Minkowski signature). For the D4 brane $E_{n_i}$ are positive solutions to
\bea
\t G^{\mu\nu} n_{\mu} n_{\nu} &=& 0
\eea
where $n_t = E_{n_i}$. We use the index notation $M = (\mu,5)$ and $\mu = (t,i)$ for $i=1,\cdots,4$. Since
\bea
g^{MN} n_M n_N &=& \t G^{\mu\nu} \(n_{\mu} - v_{\mu} n_5\)\(n_{\nu} - v_{\nu} n_5\) + \frac{1}{R^2} {n_5}^2
\eea
we see that
\bea
g^{MN} n_M n_N &=& \t G^{\mu\nu} n_{\mu} n_{\nu}
\eea
only for vanishing KK momentum $n_5 = 0$.

If we pick factors in $\Z_{osc}^{M5}$ with $n_i = 0$ and perform the product over all non-vanishing Kaluza-Klein momenta $n_5=1,2,3,\cdots$, then these give rise to a $\tau$-dependent factor
\bea
\frac{1}{\eta(\tau)^3 \overline{\eta(\tau)}^3}
\eea
Here the powers $3$ correspond to the six harmonics $b_{ij}$ on $T^4$, separated into selfdual and anti-selfdual parts. These harmonics are responsible for the 3-fold degeneracy of $E_{n_5}$. For the D4 brane we put $n_5 = 0$ and this $\tau$-dependent factor does not arise in $\Z_{osc}^{D4}$.

\section{The M5/D4 partition function on $T^6$}
Multiplying together the various contributions, what have found on $T^6$ is the following M5 brane partition function
\bea
\Z^{M5} &=& \Z^{M5}_{zero} \Z^{M5}_{osc}
\eea
where
\bea
\Z^{M5}_{zero} &=& \Theta\[^0_0\](-\tau)  \(\frac{i\Vol}{\sqrt{-\beta}}\)^{\frac{5}{2}}\\
\Z^{M5}_{osc} &=& \prod_{\p\neq 0} \(\frac{1 + e^{2\pi i E_{\p}^+}}{1 - e^{2\pi i E^+_{\p}}}\)^4 \(\frac{1 + e^{-2\pi i E_{\p}^-}}{1-e^{-2\pi i E^-_{\p}}}\)^4\label{Zosc}
\eea
and where $\vec{p} = (p_m) = (p_i,p_5)$ for $i=1,...,4$. For the D4 brane we have found that
\bea
\Z^{D4}_{zero} &=& \Z^{M5}_{zero}\cr
\Z^{D4}_{osc} &=& \prod_{p_i\neq 0} \(\frac{1 + e^{2\pi i E^+_{p_i}}}{1 - e^{2\pi i E^+_{p_i}}}\)^4 \(\frac{1 + e^{-2\pi i E_{p_i}^-}}{1-e^{-2\pi i E^-_{p_i}}}\)^4
\eea
What is missing so far from the D4 brane are the KK modes with $p_5 \neq 0$, to which we will turn our attention in the next section. Let us here just note that a particularly simple KK sector in the M5 brane oscillators is obtained by taking $p_i = 0$ with $p_5 \neq 0$. We note that
\bea
E^{\pm}_{p_i = 0, p_5} &=& \pm \sqrt{-\beta}\sqrt{G^{55}}|p_5| - V^5 p_5
\eea
and so by defining
\bea
q &=& e^{2\pi i \(\sqrt{-\beta}\sqrt{G^{55}} - V^5\)}\cr
\t q &=& e^{-2\pi i \(-\sqrt{-\beta}\sqrt{G^{55}} - V^5\)}
\eea
we find that this sector gives the following contribution
\ben
\Z^{M5}_{osc,p_i = 0} &=& \prod_{p_5 = 1}^{\infty}  \( \frac{1 + q^{p_5}}{1 - q^{p_5}} \)^8 
                                                                         \( \frac{1 + \t q^{p_5}}{1 - \t q^{p_5}} \)^8\label{SenM5}
\een

\section{Small instantons and KK modes}
Up to now we have shown that the naive D4 brane partition function on $T^5$ does not match with that for the
M5 brane on $T^6$. The zero mode's part has the precise agreement while there is a clear mismatch in the oscillator
part of the partition functions. If the proposed D4 / M5 on $S^1$ correspondence is correct, this implies that one is
missing certain degrees of freedom  from the D4 brane side. It is clear  from the previous discussion so far that
the missing part are all those spectra of KK modes along the circle direction $S^1$.  Here we would like to
show that this KK part of the partition function is precisely generated by small instantons corresponding to
 D0 branes of the type IIA string theory.

Taking our $S^1$ as an M-theory circle,   the M5 becomes D4 by the type IIA reduction and the KK circle
momenta are corresponding to  D0  branes from the view point of the type IIA theory. If  D0 branes are away
from the D4 to the five transverse directions, then their Hilbert space is not to do with that of the D4 brane.
When  D0 branes are located on top of the D4, they  can be bound to the D4 brane forming a (threshold)
bound state, which can be faithfully captured by the D4 brane worldvolume dynamics we are dealing with.
This is one of the basic construct of the D4 / M5 on $S^1$ correspondence proposed in Refs. \cite{Lambert:2010iw,
Douglas:2010iu}.
These D0 branes are captured by the well known instanton dynamics of D4 branes satisfying
the selfdual (or anti-selfdual) equation
\bea
F\ &=& (-) *_4 F
\eea
where the hodge dual $*_4$ is taken over the four spatial direction of the worldvolume
directions of D4 branes. The instanton number is counted by an integer $n_5$
\bea
n_5 &=& \frac{1}{8 \pi^2} \int {\rm tr}\,\, F \wedge F
\eea
For the $SU(N)$ part of $N$ D4 branes with $N\ge 2$,  regular
solutions of the above self dual equation can be found explicitly. It is well known, however, that
for the case of $U(1)$, the corresponding instanton solutions become singular classically.
Hence one has to introduce
some regularization in order to use the conventional techniques
including the moduli space approximation. Here we shall introduce the noncommutativity parameters
for the spatial part of the coordinates for the sake of the regularization
and then, in the end, we take the commutative limit by sending the noncommutative parameters to zero.
In this regularization, the instanton size squared turns out to be of order of the noncommutativity
parameters. Since we are interested in the small size limit, the physics of  global aspects decouples from
the local dynamics and, for instance,
the boundary condition  on their wave
function can be imposed separately afterwards.

Hence by an appropriate coordinate transformation, we make
the D4 brane worldvolume metric to the standard flat Minkowski form,
\bea
ds_5^2 &=& -dz_0^2 + dz_i dz_i
\eea
With these coordinates, the 6d metric including circle fibration now takes the form
\bea
ds^2_6 &=& ds_5^2 + R^2(dx^5 + \tilde{v}_\mu dz^\mu)^2
\eea
We shall introduce the spatial noncommutativity
\bea
[z_i, z_j] &=& i \theta_{ij}
\eea
where the noncommutativity parameters $\theta_{ij}$ can be fully parameterized
by
\bea
\theta_{ij} &=& \xi^a \eta^a_{ij} + \zeta^a \bar{\eta}^a_{ij}
\eea
where $\eta^a_{ij}$ and ${\bar\eta}_{ij}$ $a=1,2,3$ are the selfdual and anti-selfdual
't Hooft tensors respectively.

We shall focus on the Maxwell part of the D4 brane action that is given by
\ben
S &=& -\frac{1}{16\pi^2 R} \int d^5 z \, F_{\mu\nu} F^{\mu\nu} +\frac{1}{32\pi^2} \int d^5 z\, \epsilon^{\mu\nu\lambda\delta \rho}
F_{\mu\nu} F_{\lambda\delta}\, \tilde{v}_\rho
\label{noncomaction}
\een
and we are interested in the instanton solutions satisfying
\bea
F_{ij} &=& \frac{1}{2}\epsilon_{ijkl}F^{kl}
\eea
and their moduli space dynamics.
The instanton solutions of the selfdual equation and their modular space dynamics
can be systematically studied by the so called Atiyah-Drinfeld-Hitchin-Manin (ADHM)
method of the construction \cite{Atiyah:1978ri}, for which the noncommutative deformation can be
incorporated in addition \cite{Nekrasov:1998ss}.  
We shall not introduce all the details here but just necessary part for the
present purpose. The ADHM equations for selfdual instantons reads\footnote{The anti-selfdual instanton corresponds to replacing $\zeta^a$ by $\xi^a$ in the ADHM equations. Therefore, to have ADHM constructions for both instantons and anti-instantons, we need to have both $\zeta^a$ and $\xi^a$ non-vanishing.}
\bea
[Z,Z^\dagger] +[W,W^\dagger] + I I^\dagger -J^\dagger J &=& \zeta^3\equiv \zeta\cr
[Z,W]+IJ &=& \zeta^1+i\zeta^2
\eea
where $X$ and $W$ are complex $n_5 \times n_5$ matrices and $I^\dagger$ and $J$ are $N \times n_5$
complex matrices.
As before $N$ is for the $N$ parallel $D4$'s and we are here only
concerned about the $N=1$ case. We note the anti-selfdual part of the noncommutativity
parameters enters for the ADHM equations for selfdual instantons.
Since there are $U(n_5)$ gauge symmetry for the $n_5$ instanton dynamics, one has to mode
out the degrees of the above ADHM equations  by an appropriate $U(n_5)$ gauge fixing conditions.
Thus the moduli space dimension as parameterized by the matrices is given by
$4 n_5^2 + 4 n_5 N$ moded out by  $3 n_5^2$ conditions imposed by the ADHM equations together with
$n_5^2$ gauge fixing conditions. Therefore the resulting dimension becomes $4n_5 N$ or $4n_5$ for our interest of
$N=1$ case.  The moduli space metric is induced from the flat metric
\bea
ds^2 &=& {\rm tr} \left(\, dI dI^\dagger + dJ^\dagger dJ + dZ d Z^\dagger+ dW dW^\dagger \right)
\eea
by a hyper-Kahler quotient procedure \cite{Lee:2000hp}.

\subsection{One instanton dynamics}
From now on, we shall focus on $U(1)$ case with $N=1$, set $\xi_1+i \xi_2=\zeta_1+i \zeta_2=0$ and take $\zeta >0$.
Then for one instanton case with $n_5=1$, the gauge fixed  solution reads
\bea
I_{1\times 1}=\sqrt{\zeta}\,, \quad\quad J_{1\times 1}=0
\eea
while $Z_{1\times 1}$ and $W_{1\times 1}$ are not constrained at all. Thus we conclude that one instanton
moduli space is given simply  by flat $R^4$ with
\bea
ds^2_{n_5=1}&=& |dZ|^2 + |dW|^2= dX_i dX_i
\eea
where $X_i$ is the position on $R^4$.
This describes a D0 with a definite finite size  freely moving along the worldvolume
directions of the D4 brane. (\,One can show that there is a finite binding energy between $D0$ and $D4$ due the nonvanishing
noncommutativity parameter $\zeta$.)

For the details of its Hilbert space and moduli dynamics, one may be more explicit for this
simplest case. Of course via the ADHM construction, one may find explicit nonsingular one instanton
solution $A_i (\vec{z})$ with
$A_0=0$ \cite{Nekrasov:1998ss,Bak:2000ac} but we do not need here its explicit form since the translational
symmetry of the system
will be enough to show some details as will be clear below. With the choice of $A_0=0$, the configuration
$A_i(\vec{z}-\vec{X})$ is explicitly describing the moduli space $R^4$.  Here $X_i$ are the so-called collective
coordinates for the moduli (solution) space interpreted as a position of D0 along the worldvolume
directions of the D4. In general the zero mode that is the variation of the solution
tangent to the moduli space requires an extra gauge transformation
\bea
\delta_j A_i &=& \frac{\partial}{\partial X_j} A_i + \frac{\partial}{\partial z_i} \lambda_j
\eea
where $\lambda_j$ will be chosen such that the constraints $\frac{\partial}{\partial {z_i}} \delta_j A_i=0 $ hold, which means that the zero modes are orthogonal to gauge transformation modes $\delta A_i = D_i \Lambda$. For the present one instanton case, the constraints are trivially solved with $\lambda_j=0$ by assuming the gauge fixing condition $\partial_i A_i = 0$. Thus for the explicit moduli space dynamics, we insert the ansatz
\bea
A_i(\vec{z}-\vec{X}(t)) \,, \quad\quad A_0=0
\eea
to the action (\ref{noncomaction}) in order to obtain the low energy effective action for the
one-instanton dynamics. This will be valid up to quadratic order of the velocity $\dot{X}_i$.
For the evaluation we further note
\bea
&& \frac{1}{16\pi^2} \int d^4 z F_{ij} F_{ij}=n_5=1\\
&&  \frac{1}{16\pi^2} \int d^4 z F_{ij} F_{im}=\frac{n_5}{4}\delta_{jm}
\eea
where for the latter we have used the translational symmetry of the one-instanton solution.
The resulting effective Lagrangian for the one-instanton dynamics reads
\bea
L^{n_5=1}_{eff} &=& -\frac{n_5}{R}+ n_5 \, \tilde{v}_0 \,
+ \frac{n_5}{2R} \dot{X}_i \dot{X}_i + n_5 \, \tilde{v}_i \dot{X}^i
\eea
For anti-instantons we instead find (with $n_5 = -1$)
\bea
L^{n_5=-1}_{eff} &=& \frac{n_5}{R} + n_5 \, \tilde{v}_0 \,
- \frac{n_5}{2R} \dot{X}_i \dot{X}_i + n_5 \, \tilde{v}_i \dot{X}^i
\eea
The canonical momentum has an extra contribution from the first-order time derivative term,
\bea
p_i &=& \pm \frac{n_5}{R} \dot{X}_i + n_5  \, \tilde{v}_i
\eea
with upper sign for instanton and lower sign for anti-instanton respectively. The Hamiltonian becomes
\bea
H &=& \pm \frac{n_5}{R} - n_5 \, \tilde{v}_0 \pm \frac{R}{2 n_5 } \left(
p_i - n_5 \tilde{v}_i
\right) \left(
p_i - n_5 \tilde{v}_i
\right)
\eea
Here $\pm \frac{n_5}{R}$ is the mass of the nonrelativistic motion of an instanton and of an anti-instanton respectively. We note that the sign is precisely such that this mass is always positive. One then recognizes that this is the nonrelativistic version of the relativistic counterpart
\ben
-(p_0 - n_5 \tilde{v}_0 )^2 + \left(
p_i - n_5 \tilde{v}_i
\right)^2 + \frac{n_5^2}{R^2} &=& 0\label{eom}
\een
in Minkowski signature if we make the following identifications
\bea
H = - p^-_0 > 0
\eea
for the instanton and the anti-instanton respectively. Here $p_0^{\pm}$ denote the positive and negative roots to the quadratic equation of motion (\ref{eom}) (for a sufficiently small $\t v_0$). Since one has the 5d Poincar\'e symmetry of  D4 in the commutative limit, one may
argue that the above should be the precise spectrum; Its 6d version takes the form
\bea
g^{MN}p_M p_N &=& 0
\eea
with the instanton number $n_5$ is identified with the KK momentum $p_5$ along
the circle direction. Thus we conclude that the required spectra for  single KK mode are precisely
reproduced from the single instanton and anti-instanton dynamics respectively.

The supersymmetric extension of this analysis for the one-instanton case is rather
straightforward. We shall describe simply the result here. The one-instanton is half BPS
(preserving half of the 16
supersymmetries)  and the 8 fermionic zero modes form a multiplet consisting of
8 bosonic and 8 fermionic states. This agrees with the ground state degeneracy of one-D0 and one-D4
bound states.  Among the 8 bosonic states, $\bf 5$ correspond to the KK parts of 5 scalar
while the remaining $\bf 3$ are the KK modes from the selfdual two form gauge field.  The fermionic
part are also matching precisely with the KK modes from $\bf 8$ fermionic degrees.
Thus we conclude that the dynamics of one instanton sector is precisely matching with that
of the one KK mode of the M-theory circle.

\subsection{ Multi instanton dynamics}

We now turn to the cases of higher instanton numbers. The case of two $U(1)$ noncommutative instantons
has been analyzed in \cite{Lee:2000hp} which leads to the metric on the 8 dimensional moduli space
\bea
ds_{n_5=2}^2 &=& ds^2_{\rm com} + ds^2_{\rm rel}
\eea
where the center of mass part is $\mb{R}^4$ with metric
\bea
 ds^2_{\rm com} &=& dX_i d X_i
\eea
while the relative part of the metric is given by the Eguchi-Hanson metric \cite{Eguchi:1978gw}
\bea
 ds^2_{\rm rel} &=& \frac{r^2}{\sqrt{r^4+ {4 \zeta^2}}} \left(
dr^2 + \frac{r^2}{4} \sigma_3^2
\right) + \frac{1}{4} \sqrt{r^4+ {4 \zeta^2}} \,  \left(
 \sigma_1^2 + \sigma_2^2
\right)
\eea
In this metric $\sigma_a$ are the standard $SU(2)$ left invariant one forms given by
\bea
\sigma_1 &=& -\sin \psi d \theta + \cos \psi \sin \theta d \varphi \\
\sigma_2 &=& \cos \psi d \theta + \sin \psi \sin \theta d \varphi \\
\sigma_3 &=&  d\psi  + \cos \theta d \varphi
\eea
and the angular variables are ranged over
\bea
0 \le \theta \le \pi \,, \quad \quad 0 \le \varphi \le 2\pi\ \,, \quad \quad 0 \le \psi \le 2\pi
\eea
due to these angular ranges, the geometry involves $S^3/\mb{Z}_2$ whereas a full $S^3$ requires the range
$0 \le \psi \le 4\pi$.
As expected, this metric is nonsingular at $r=0$. In the commutative limit of $\zeta\rightarrow 0$, one may ignore any interactions
between instantons completely leading to the metric  for
\bea
\mb{R}^4 \times \mb{R}^4/\mb{Z}_2
\eea
or
\bea
T^4 \times T^4/\mb{Z}_2
\eea
including the global boundary conditions,
where $\mb{Z}_2$ is interpreted as the permutation symmetry of identical particles. Thus the Hilbert space for  two KK modes
is realized by the two instanton dynamics.

Their quantum mechanical Hamiltonian is given by the Laplacian operators of the geometry acting on the
wave functions with internal structure characterized by $n$-forms. It is well known that the Eguchi-Hanson metric
allows a unique normalizable selfdual harmonic two-form  state
\bea
\omega_2 &=& -\frac{2 r^3}{(r^4+4 \zeta^2)^{\frac{3}{2}}} dr \wedge  \sigma_3 +\frac{1}{(r^4+4 \zeta^2)^{\frac{1}{2}}}
\sigma_1 \wedge \sigma_2
\eea
This is interpreted as a threshold bound state two $U(1)$ instantons. As noticed in \cite{Lee:2000hp}, this threshold
bound state corresponds to the state of the charge two ($n_5=2$) KK mode.

For the maximally supersymmetric case, the counting of relevant two-instanton states  goes as follows.
As explained each D0 brane has  $2^4$ states, eight bosonic and eight fermionic.
Now for two D0 branes, considering them as identical particles, the exchange symmetry has to be incorporated
in this counting of internal states. There are $\frac{1}{2}\times 8 \times 9$ possible states with  both D0 branes in the
bosonic states. There are also $\frac{1}{2}\times 8 \times 7$ states with both D0 branes in (different) fermionic states
(due to the exclusion principle).  The remaining possibilities include $8 \times 8$ states with one in a bosonic state and
one in a fermionic state. Hence the two (separate) D0 branes involve in total $128$ states.
But as just described above, two D0 branes can also form a threshold bound state and be bound to the
D4 brane, which is then identified as a single $n_5=2$ KK mode. This involves $16$ states as a single entity.
Thus the total number of states with $n_5=2$ is counted as $d_2 = 16 + 128= 144$. 

Threshold bound states for $n_5 > 2$ have not yet been constructed. However their existence was proven in  \cite{Kim:2011mv}. For $n_5 = 3$ there is thus a threshold bound state which corresponds to a single instanton particle carrying instanton number $n_5 = 3$ and which has $16$ internal states (eight bosonic plus eight fermionic). We can also have two separate instanton particles, one with instanton number $n_5 = 1$ and the other being a treshold bound state with instanton number $n_5 = 2$. Since these two instanton particles are distinct they contribute with $16 \times 16 = 256$ internal states. Finally we can have three separate instantons each carrying instanton number $n_5 = 1$. As each of these instantons can be either bosonic or fermionic they add up to $120 + 288 + 224 + 56 = 688$ internal states. (The four terms correspond to instanton types $BBB$, $BBF$, $BFF$, $FFF$ where $B$ and $F$ refers to $8$-bosonic and $8$-fermionic states.) Thus for $n_5 = 3$ we have in total $d_3 = 16 + 256 + 688 = 960$ states. 
In general we get the total number of degrees of freedom $d_{n_5}$ at a given instanton number $n_5$ from the following generating partition function as
\ben
2^8 \sum^\infty_{n=0} d_n \, q^n &=& 2^8 \prod^\infty_{n_5=1} \left(\frac{1+q^{n_5}}{1-q^{n_5}}\right)^8\label{SenD4}
\een
where the overall $2^8$ is the counting of states for D4 brane ground states. This formula
can be derived from the counting of BPS states in string theory and using some chains of
$U$ dualities \cite{Sen:1995hb}, \cite{Kim:2011mv}.
We have explicitly checked that the coefficient $d_1$, $d_2$ and $d_3$ are correctly generated by this generating function.

The counting is precisely in parallel with the required KK states which is the missing part of M5 brane partition function
in the naive D4 brane computation without incorporating instanton contributions. We note that $\Z^{M5}_{osc,p_i=0}$ as obtained in eq (\ref{SenM5}) contains as one factor the generating function of instantons (\ref{SenD4}). The second factor in eq (\ref{SenM5}) which involves a product over $\t q$ shall be interpreted as the contribution from anti-instantons.  

Instantons are BPS configurations, which means that static instantons do not interact among themselves. Static anti-instantons likewise do not interact among themselves. However an instanton interacts with an anti-instanton since such a configuration is not BPS. Also if we give some small momenta to instantons, they start to interact and feel the metric of the moduli space. However as the instanton and anti-instanton sizes become small in the commutative limit, the moduli space becomes flat and these become non-interacting particles. In the same fashion instanton and anti-instanton interactions disappear in this limit. Their contribution to the D4 brane partition function then factorizes. First from the instantons we have the factor
\bea
\Z^{inst} &=& \prod_{n_i \in \mb{Z}} \Z^{inst}_{n_i}
\eea
where $n_i$ labels momenta of  an instanton or a threshold-bound state of instantons which we call a KK particle with $n_5 =1$ or with some nonvanishing
 $n_5$ respectively. Using the counting of degrees of freedom as obtained above up to instanton number $n_5 = 3$, we have the following contribution from KK particles with a specific momentum $n_i$ to the D4 brane partition function
\bea 
\Z^{inst}_{n_i} &=& 1+ 16 \, e^{2\pi i E^+_{n_i,1}}+16\,  e^{2\pi i E^+_{n_i,2}}+128\, e^{4\pi i E^+_{n_i,1}}\cr
&& + 16 e^{2\pi i E^+_{n_i,3}} + 256 e^{2\pi i \(E_{n_i,1}^+ + E_{n_i,2}^+\)} + 688 e^{6\pi i E^+_{n_i,1}}\cr
&& +  \dots
\eea
where $E^+_{n_i,n_5}$ refers to $-p_0^- >0$ where $p_0^{\pm}$ are the two roots to the equation (\ref{eom}).  Second from the anti-instantons we have the factor
\bea
\Z^{anti-inst} &=& \prod_{n_i} \Z^{anti-inst}_{n_i}
\eea
As we have seen, the anti-instanton is associated with $E^+_{n_i,n_5} = -p_0^- > 0$ for $n_5$ negative. We then find that 
\bea
\Z^{anti-inst}_{n_i} &=& 1+ 16 \, e^{2\pi i E^+_{n_i,-1}}+16\,  e^{2\pi i E^+_{n_i,-2}}+128\, e^{4\pi i E^+_{n_i,-1}}\cr
&& + 16 e^{2\pi i E^+_{n_i,-3}} + 256 e^{2\pi i \(E_{n_i,-1}^+ + E_{n_i,-2}^+\)} + 688 e^{6\pi i E^+_{n_i,-1}} \cr
&&+  \dots
\eea
We now claim that $n_5 \neq 0$ contributions to the M5 brane partition function contain generating functions for these instanton and anti-instanton threshold bound states. More precisely, we claim that 
\bea
\Z^{inst}_{n_i}  &=& \prod_{n_5 = 1}^{\infty} \(\frac{1+ e^{2\pi i E_{n_i,n_5}^+}}{1 - e^{2\pi i E^+_{n_i,n_5}}}\)^8\cr
\Z^{anti-inst}_{n_i} &=& \prod_{n_5 = -\infty}^{-1} \(\frac{1 + e^{2\pi i E_{n_i,n_5}^+}}{1-e^{2\pi i E^+_{n_i,n_5}}}\)^8
\eea
It is rather easy to check explicitly that these generating functions reproduce the correct number of degrees of freedom for each threshold bound state up to total instanton number $n_5 = 3$. By noting the relation
\bea
E^+_{n_i,n_5} &=& -E^-_{-n_i,-n_5}
\eea
it is easy to see that we have
\bea
\Z^{M5}(T^6) &=& Z^{D4}(T^5) Z^{inst}(T^5) Z^{anti-inst}(T^5)
\eea
where the various factors on the right-hand side correspond to instanton numbers $n_5 = 0$, $n_5 > 0$ and $n_5 < 0$ respectively. We may also phrase this relation as
\bea
\Z^{M5}(T^6) &=& \lim_{\zeta,\xi \rightarrow 0} \Z^{D4}_{\zeta,\xi}(T^5)
\eea
since the non-commutative D4 already has those instanton sectors in the theory so they shall not be supplemented. 

There is now little doubt that the M5 brane is really the same thing as (noncommutative) D4 in the above sense. In particular the M5 brane partition function encodes the number of internal degrees of freedom of each D0 brane threshold bound state.

One final note is the fact that the zero mode part of the partition function include also so called large instanton
configurations which are satisfying the same (anti) selfdual equation of instantons.  However as we showed explicitly, these large
instantons are nothing to do with D0 branes. Namely, without involving any D0 branes, we have  shown that
the zero mode part of D4 and M5 has already an agreement with each other.

Since the essential  properties of small instanton contribution we are using are only local ones, one expects that the above discussion
can be extended to the case of general five manifold with circle fibration. But the relevant ADHM construction and the corresponding
regularization is lacking currently. We shall leave this issue for future investigations.

\section{Singular fibration}
So far we have considered circle fibers with constant radius $R$. We may also consider situations where $R$ depends on the D4 brane worldvolume coordinate. At some points one may also allow $R=0$ in which case we have a singular fibration. From the M5 brane worldvolume point of view such singular fibrations are smooth and pose no further problems in the computation of the partition function, but from the D4 brane point of view the computation of the partition function becomes much more difficult when $R$ is not constant. Also we may need to add new degrees of freedom at the locus of the singularity $R=0$. As a singular fibration we may consider the M5 brane worldvolume on the form $W = TN \times T^2$ with $TN$ being a Taub-NUT space that supports a single harmonic selfdual two-form $\Omega^+$. Such a fibration has been considered also in \cite{Ohlsson:2012yn}, \cite{Witten:2009at}. Then the zero mode part of the M5 brane partition function will involve $\Theta(\tau)$ where $\tau$ is the period matrix on $T^2$. There is no fully modular invariant theta function except for $\Theta\[^{\frac{1}{2}}_{\frac{1}{2}}\](\tau)$ that vanishes\footnote{That is not so if we couple M5 brane to a background $C$ field though.}, so the M5 brane partition function will depend on the choice of spin structure. As symplectic basis of harmonic three-forms on $W$ we take
\bea
\alpha &=& \Omega^+ \wedge dx\cr
\beta &=& \Omega^+ \wedge dy
\eea
and the period matrix of $W$ becomes the $\tau$ parameter of $T^2$. On the five-dimensional base-manifold $\mb{R}^3 \times T^2$ we have sYM theory with coupling constant $g^2 \sim R$. It appears that the zero mode contribution from sYM can not give us a $\tau$ dependence of the partition function because the only harmonic two-form is the one on $T^2$ itself, while to have a $\tau$ dependence we would need a harmonic two-form that has only one component on $T^2$. Such a harmonic two-form is not normalizable on $\mb{R}^3 \times T^2$ and must be excluded. Instead we have to add a chiral scalar action localized at the singularity $\{0\}\times T^2$ of $TN \times T^2$. The chiral scalar of course gives us a partition function $\Theta(\tau)$ on $T^2$.

Another way to see this is by studying the gauge anomaly cancelation. The graviphoton term
\bea
S_{sYM} &=& \frac{1}{8\pi^2}\int_{\mb{R}^3 \times T^2}  A \wedge dA \wedge w
\eea
with
\bea
w &=& -\frac{1}{2r^3} \epsilon_{ijk} x^i dx^j \wedge dx^k
\eea
has the anomalous gauge variation
\bea
\delta S_{sYM} &=& \frac{1}{8\pi^2} \int_{T^2 \times R_+} d\lambda \wedge F \int_{S^2} w\cr
&=& \frac{1}{4\pi} \int_{T^2} \lambda F
\eea
under the variation
\bea
\delta A &=& d\lambda
\eea
This gauge anomaly can be canceled by a gauged scalar field theory
\bea
S_{WZW} &=& \frac{1}{8\pi}\int_{T^2} \(|d\phi + A|^2 + 2\phi F\)
\eea
supported on $\{0\} \times T^2$ whose gauge variation is
\bea
\delta S_{WZW} &=& -\frac{1}{4\pi} \int_{T^2} \lambda F
\eea
under
\bea
\delta \phi &=& -\lambda\cr
\delta A &=& d\lambda
\eea
Moreover, $2\pi$ periodicity of the holonomy $\int A$ implies that $\phi$ is also $2\pi$ periodic in order to make $d\phi + A$ a gauge invariant quantity. Now this together with the normalization we found for $S_{WZW}$ tells us that we have a coupling constant which corresponds to the free fermion radius. This action corresponds to Chern-Simons theory in three dimensions at level $k=1$. Such a Chern-Simons theory may be derived from seven dimensions by expanding
\bea
C &=& \Omega^+ \wedge A
\eea
We then get
\bea
\frac{k}{4\pi}\int_{\mb{R}\times W} C \wedge dC &=& \frac{k}{4\pi}\int_{TN} \Omega^+ \wedge \Omega^+ \int_{\mb{R}\times T^2} A \wedge dA
\eea
The selfdual harmonic two-form was found in \cite{Gauntlett:1996cw}, \cite{Lee:1996if}. We will normalize it so that $\Omega^+ = \frac{1}{4\pi}\omega$ where
\bea
\omega &=& \frac{r}{r+1} \sigma_1 \wedge \sigma_2 + \frac{1}{(r+1)^2} dr \wedge \sigma_3
\eea
using the notation and conventions of \cite{Gauntlett:1996cw}. Then we have
\bea
\int_c \Omega^+ &=& 1\cr
\int \Omega^+ \wedge \Omega^+ &=& 1
\eea
where the three-cycle $c$ is spanning the $r,\psi$ plane where the coordinate range is $\psi \in [0,4\pi]$. We then find that $A$ is a connection one-form whose holonomy is $2\pi$ periodic, induced from $C$, and we find that
\bea
\frac{k}{4\pi} \int A \wedge dA
\eea
with standard normalization, induced from Chern-Simons action for $C$.

Another generalization we did not consider in this paper is to include six-manifolds which have no one-cycles at all. As we can see from our Appendix $A$, once we have a one-cycle, there is a natural choice of symplectic basis of three-forms and corresponding three-cycles. If we have no one-cycle in our six-manifold we may of course still find a symplectic basis though the method in Appendix $A$ does not apply to such cases. For example we may consider the case when the six-manifold is $W = S^3 \times S^3$. On $S^3$ we do not have a one-cycle even though it can be viewed as a Hopf bundle over a base manifold $S^2$. Nevertheless we can define a symplectic basis for this situation by taking as basis three-forms certain rescaled volume forms on each of the two $S^3$ respectively. One may also consider a situation where the two Hopf fibers make up a skew torus, in which case the period matrix $\tau$ (which is just a complex number) can take any value in the upper half-plane. Also we have been ignorant about curvature corrections and essentially our presentation in the present paper is correct only for a flat $T^6$. On curved space we need to add a curvature term for the scalar fields to maintain conformal invariance. Moreover superconformal invariance puts severe constraints on the possible six-manifolds one may consider \cite{Linander:2011jy}. However one may circumvent this constraint from superconformal invariance by considering a situation where one has a background flux on $S^3$ which spontaneously breaks conformal symmetry, while preserving maximal supersymmetry. Such a situation was considered in \cite{Gustavsson:2012ei} where it was shown that the zero mode parts of the M5 brane partition function on $T^3 \times S^3$ matches with the corresponding zero mode part of the D4 brane partition function on $T^3\times S^2$ respectively. It would be interesting to extend this analysis to the oscillator modes.

\section*{Acknowledgement}
DB would like to thank Seok Kim for helpful discussions. This work was
supported in part by NRF SRC-CQUeST-2005-0049409 and  NRF Mid-career Researcher
Program 2011-0013228.

\newpage
\appendix
\section{Period matrix on $S^1 \times M_5$}\label{period1}
We assume a six-manifold on the form $M_6 = S^1 \times M_5$. On the space of harmonic two-forms $\Omega^i$ on $M_5$, we define a metric
\bea
\G^{ij} &=& \int \Omega^i \wedge * \Omega^j
\eea
where $*$ denotes the Hodge star on $M_5$. We define a basis of dual harmonic three-forms
\bea
\t \Omega_i &=& 2\pi \G_{ij} * \Omega^j
\eea
where $\G_{ij}$ is the inverse of $\G^{ij}$. Then we have a symplectic basis
\bea
\alpha_i &=& \t \Omega_i\cr
\beta^i &=& \Omega^i \wedge dt
\eea
We assume $\int dt = 2\pi$ and that $\int_{C_j} \Omega^i = \delta_j^i$ when integrated over dual two-cycle. The $\alpha_i$ and $\beta^i$ then have periods $2\pi$ when integrated over their dual three-cycles in $M_6$. We also have the symplectic property
\bea
\int \alpha_i \wedge \beta^j &=& 4\pi^2 \delta_i^j
\eea
One may object that we should not need a metric to define a symplectic basis. As we will see in the next section, if $M_6 = T^2 \times M_4$ then this symplectic basis can be rewritten in terms of the intersection form on $M_4$ and the metric does not enter the definition of the symplectic basis.

We define the period matrix $\tau_{ij}$ and its conjugate $\bar{\tau}_{ij}$ by requiring that
\bea
\omega_i &=& \alpha_i + \tau_{ij} \beta^j \cr
\bar{\omega}_i &=& \alpha_i + \bar{\tau}_{ij} \beta^j
\eea
are selfdual and antiselfdual respectively with respect to the six-dimensional Hodge star that we define with respect to the six-dimensional metric tensor
\bea
g^{MN} &=& \(\begin{array}{cc}
\frac{1}{\beta} & \frac{1}{\beta} V^n\\
\frac{1}{\beta} V^m & G^{mn} + \frac{1}{\beta} V^m V^n
\end{array}\)
\eea
We will define this six-dimensional Hodge-star when acting on three-forms as
\bea
*_6 dx^M \wedge dx^N \wedge dx^P &=& \frac{1}{6}\sqrt{-g} \epsilon^{MNP}{}_{M'N'P'} dx^{M'} \wedge dx^{N'} \wedge dx^{P'}
\eea
This definition is related to the conventional Hodge-star as $*_6 = \frac{\sqrt{-\beta}}{\sqrt{|\beta|}} *_{6,conventional}$. Our Hodge-star depends holomorphically on $\beta$ which we may promote into a holomorphic coordinate. Our Hodge star now squares according to
\bea
*_6 *_6 &=& 1
\eea
when acting on three-forms in six dimensions, and so we shall by selfduality/antiselfduality mean that
\bea
*_6 \omega &=& \omega\cr
*_6 \bar{\omega} &=& -\bar{\omega}
\eea
It turns out that the solution to these conditions can be expressed like
\bea
\tau_{ij} &=& -(\tau_1)_{ij} - (\tau_2)_{ij}
\eea
where
\bea
\alpha_i &=& (\tau_1)_{ij} \beta^j + (\tau_2)_{ij} *_6 \beta^j
\eea
If we expand
\bea
*_6 \beta^i &=& C^{ij} \alpha_j + A^i{}_j \beta^j
\eea
then
\bea
X_{ij} &=& (C^{-1})_{ik} A^k{}_j\cr
Y_{ij} &=& (C^{-1})_{ij}
\eea
From the symplectic properties, it follows that
\bea
C^{ij} &=& \frac{1}{4\pi^2} \int *_6 \beta^i \wedge \beta^j\cr
A^i{}_j &=& -\frac{1}{4\pi^2} \int *_6 \beta^i \wedge \alpha_j
\eea
By brute force computation we get
\bea
\int *_6 \beta^i \wedge \beta^j &=& \frac{2\pi}{\sqrt{-\beta}} \G^{ij}\cr
\int *_6 \beta^i \wedge \alpha_j &=& \frac{2\pi}{\sqrt{-\beta}} \G_{jk} 2\pi \L^{ki}
\eea
Here we define
\bea
\K^i{}_j &=& \frac{1}{2}\int d^5 x \sqrt{G}\Omega^i_{mn} \t \Omega^{mnp}_j V_p\cr
\L^{ij} &=& \frac{1}{4}\int d^5 x G \epsilon^{mnpqr} \Omega^i_{mn} \Omega^j_{pq} V_r
\eea
and we have the relation
\bea
\K^{ij} &=& 2\pi \L^{ij}
\eea
however the latter form $\L^{ij}$ shows manifest symmetry in $ij$. Then we get the period matrix as
\bea
\tau_{ij} &=& 2\pi \(-\L_{ij} + \sqrt{-\beta} \G_{ij}\)
\eea

\section{Period matrix on $T^2\times M_4$}\label{period2}
For simplicity we assume that $M_4$ has no one-cycles. Then the harmonic two-forms on $M_4$ must be the same as the harmonic two-forms on $M_5 = S^1 \times M_4$. Let us here denote these harmonic two-forms as
\bea
\Omega^I = \frac{1}{2}\Omega^I_{mn} dx^m \wedge dx^n = \frac{1}{2}\Omega^I_{ij} dx^i \wedge dx^j
\eea
where $x^m=(x,x^i)$ for $i =1,2,3,4$ and $x$ denotes the coordinate on the spatial $S^1$ in $M_5$. We define the intersection form and the metric on $H^2(M_4,\mb{Z})$ as
\bea
Q^{IJ} &=& \int_{M_4} \Omega^I \wedge \Omega^J\cr
G^{IJ} &=& \int_{M_4} \Omega^I \wedge *_4 \Omega^J
\eea
We then find that
\bea
*_4 \Omega^I &=& X^I{}_J \Omega^J
\eea
where
\bea
X^I{}_J &=& G^{IK} (Q^{-1})_{KJ}
\eea
By $*_4 *_4 = 1$ when acting on two-forms, we get
\bea
G^{IK} (Q^{-1})_{KJ} &=& Q^{IK} G_{KJ}
\eea
We define a symplectic basis on $M_6$ as
\bea
\alpha_I &=& (Q^{-1})_{IJ} \Omega^J \wedge dx\cr
\beta^I &=& \Omega^I \wedge dt
\eea
and indeed this is consistent with our choice of symplectic basis on $S^1 \times M_5$ in our previous section. To see this we first notice that
\bea
\G^{IJ} &=& 2\pi R G^{IJ}
\eea
and then
\bea
2\pi \int_{M_5} *_5 \Omega^K \wedge \Omega^L &=& \G^{KI} (Q^{-1})_{IJ} \int_{M_5} \Omega^J \wedge \Omega^L \wedge dx
\eea
Since $\Omega^I$ is an arbitrary element of $H^2(M_4,\mb{Z})$ we conclude that
\bea
2\pi \G_{IK} *_5 \Omega^K &=& (Q^{-1})_{IJ} \Omega^J \wedge dx
\eea
We define the period matrix as
\bea
\tau_{IJ} &=& -(\tau_1)_{IJ} - (\tau_2)_{IJ}
\eea
where
\bea
\alpha_I &=& (\tau_1)_{IJ} \beta^J + Y_{IJ} *_6 \beta^J
\eea
We have
\bea
*_6 \beta^I &=& *_4 \Omega^I \wedge *_2 dt
\eea
and we have
\bea
*_4 \Omega^I &=& G^{IK} (Q^{-1})_{KJ} \Omega^J\cr
*_2 dt &=& \frac{\sqrt{\gamma}}{\sqrt{-\beta}}\(-dx+V^x dt\)
\eea
We define the two-dimensional period matrix as
\bea
\tau &=& -\tau_1 - \tau_2
\eea
where
\bea
\alpha &=& \tau_1 \beta + \tau_2 *_2 \beta
\eea
and where
\bea
\beta &=& dt\cr
\alpha &=& dx
\eea
From this definition we get
\bea
\tau &=& -V^x + \frac{\sqrt{-\beta}}{\sqrt{\gamma}}
\eea
We get the six-dimensional period matrix as
\bea
\tau_{IJ} &=& \tau_1 (Q^{-1})_{IJ} + \tau_2 G_{IJ}
\eea

\section{Holomorphic factorization}
The Hamiltonian is
\bea
H &=& -\tau_2 \(n^2 + \frac{1}{4}m^2\) - \tau_1 n m\cr
&=& \frac{1}{2} \(\(\frac{m}{2}+n\)^2 \tau - \(\frac{m}{2}-n\)^2 \bar{\tau}\)
\eea
The partition function is
\bea
\sum e^{2\pi i H} &=& \sum e^{i\pi \tau \(n+\frac{m}{2}\)^2} e^{-i\pi \bar{\tau} \(n-\frac{m}{2}\)^2}
\eea
Following page 115 in \cite{Ginsparg:1988ui} we define
\bea
q &=& e^{2\pi i \tau}\cr
\bar{q} &=& e^{-2\pi i \bar{\tau}}
\eea
and compute
\bea
\sum |\Theta\[^{\alpha}_{\beta}\](-\tau)|^2 &=& \sum \(q^{\frac{1}{2}n^2} \bar{q}^{\frac{1}{2}m^2} + q^{\frac{1}{2}\(n+\frac{1}{2}\)^2} \bar{q}^{\frac{1}{2}\(m+\frac{1}{2}\)^2}\)\frac{1}{2}\(1+(-1)^{n+m}\)
\eea
As the summand vanishes unless $n+m = 2p$ is even, we can substitute
\bea
m &=& 2p - n
\eea
and we get a sum over $n,p$
\bea
\sum \(q^{\frac{1}{2}n^2} \bar{q}^{\frac{1}{2}(2p-n)^2} + q^{\frac{1}{2}\(n+\frac{1}{2}\)^2} \bar{q}^{\frac{1}{2}\(2p-n+\frac{1}{2}\)^2}\)
\eea
We substitute
\bea
n &=& p-q
\eea
and we get
\bea
&&\sum \(q^{\frac{1}{2}(p-q)^2} \bar{q}^{\frac{1}{2}(p+q)^2} + q^{\frac{1}{2}\(p-q+\frac{1}{2}\)^2} \bar{q}^{\frac{1}{2}\(p+q+\frac{1}{2}\)^2}\)\cr
&=& \sum q^{\frac{1}{2}\(\frac{m}{2}-q\)^2} \bar{q}^{\frac{1}{2}\(\frac{m}{2}+q\)^2}\cr
&=& \sum e^{2\pi i H}
\eea
since in the second line we may break the sum over $m$ into sum over $m=2p$ and a sum over $m = 2p+1$.

\end{document}